\documentclass{iucrjournals}

\usepackage{siunitx}
\usepackage{textgreek}

\title{High frame rate RIXS spectroscopy using a JUNGFRAU detector with an iLGAD sensor}

\author[a]{Nuno Duarte\IUCrCemaillink{nuno.duarte@xfel.eu}\IUCrOrcidlink{0000-0001-7019-1314}\IUCrAufn{Present address: CERN, 1211 Geneva 23, Switzerland}}
\author[a]{Lo\"{i}c~Le~Guyader\IUCrCemaillink{loic.le.guyader@xfel.eu}\IUCrOrcidlink{0000-0002-5731-3724}}
\author[b]{Viktoria~Hinger\IUCrCemaillink{viktoria.hinger@psi.ch}\IUCrOrcidlink{0000-0002-2616-4084}}
\author[a]{Marco~Ramilli\IUCrCemaillink{marco.ramilli@xfel.eu}\IUCrOrcidlink{0009-0008-8121-2027}}
\author[a]{Justine~Schlappa\IUCrCemaillink{justine.schlappa@xfel.eu}\IUCrOrcidlink{0000-0003-3343-9119}}
\author[a]{Robert~Carley\IUCrCemaillink{robert.carley@xfel.eu}\IUCrOrcidlink{0000-0002-0015-7089}}
\author[b]{Maria~Carulla\IUCrCemaillink{maria.carulla@psi.ch}}
\author[a]{Yi-Ping~Chang\IUCrCemaillink{yi-ping.chang@xfel.eu}\IUCrOrcidlink{0000-0002-2967-8910}}
\author[a]{Devesh~Chopra\IUCrCemaillink{devesh.chopra@xfel.eu}\IUCrOrcidlink{0009-0006-6987-2044}}
\author[a]{Natalia~Gerasimova\IUCrCemaillink{natalia.gerasimova@xfel.eu}\IUCrOrcidlink{0009-0000-0486-367X}}
\author[b]{Michael~Grimes\IUCrCemaillink{michael.grimes@psi.ch}}
\author[b]{Aldo~Mozzanica\IUCrCemaillink{aldo.mozzanica@psi.ch}\IUCrOrcidlink{0000-0001-8372-815X}}
\author[a]{Sergii~Parchenko\IUCrCemaillink{sergii.parchenko@xfel.eu}\IUCrOrcidlink{0000-0002-4320-4957}}
\author[a]{Le~Phuong~Hoang\IUCrCemaillink{le.phuong.hoang@xfel.eu}\IUCrOrcidlink{0000-0002-3083-815X}}
\author[b]{Urs~Staub\IUCrCemaillink{urs.staub@psi.ch}\IUCrOrcidlink{0000-0003-2035-3367}}
\author[b]{Jiaguo~Zhang\IUCrCemaillink{jiaguo.zhang@psi.ch}\IUCrOrcidlink{0000-0001-7446-210X}}
\author[b]{Anna~Bergamaschi\IUCrCemaillink{anna.bergamaschi@psi.ch}\IUCrOrcidlink{0000-0001-7817-6493}}
\author[a]{Andreas~Scherz\IUCrCemaillink{andreas.scherz@xfel.eu}\IUCrOrcidlink{0000-0001-8187-7178}}
\author[b]{Bernd~Schmitt\IUCrCemaillink{bernd.schmitt@psi.ch}\IUCrOrcidlink{0000-0002-5778-0680}}
\author[a]{Monica~Turcato\IUCrCemaillink{monica.turcato@xfel.eu}}

\affil[a]{European XFEL, Holzkoppel 4, 22869 Schenefeld, Germany}
\affil[b]{Paul Scherrer Institut, Forschungsstrasse 111, 5232 Villigen PSI, Switzerland}

\begin{document} 
\maketitle 

\begin{synopsis}
We demonstrate time-resolved RIXS measurements in the soft X-ray regime obtained at an unprecedented frame rate using a JUNGFRAU detector with an iLGAD sensor at the hRIXS spectrometer of the European XFEL. These results open up new possibilities for studying fast processes in matter through time-resolved RIXS with excellent energy and time resolution.
\end{synopsis}

\begin{abstract}
Resonant inelastic X-ray scattering (RIXS) is a powerful photon-in, photon-out spectroscopy technique for probing electronic, magnetic, and lattice excitations in matter. Time-resolved RIXS extends this capability through a stroboscopic optical pump-probe scheme to characterize the time evolution of the photoexcitation and subsequent relaxation dynamics of a sample. This technique is, however, extremely photon-hungry, requiring high-repetition-rate and intense X-ray facilities. The Heisenberg RIXS (hRIXS) spectrometer at the Spectroscopy and Coherent Scattering (SCS) instrument of the European X-ray Free-Electron Laser (EuXFEL) is designed to exploit high-repetition-rates, while maintaining optimal time and energy resolution. In this work, we demonstrate the successful deployment of a JUNGFRAU detector equipped with an inverse Low Gain Avalanche Diode (iLGAD) sensor for time-resolved RIXS studies in the soft X-ray range, using the hRIXS spectrometer. A spatial resolution of $19.71\pm0.7~\si{\micro\meter}$ and a resolving power exceeding 10,000 were achieved at an unprecedented frame rate of \SI{47}{\kilo\hertz}. Intra-train resolved data measured with a high FEL peak fluence of \SI{1.8}{\milli\joule/\centi\meter\squared} for a \SI{928.5}{\electronvolt} photon energy and \SI{1.1}{\mega\hertz} repetition rate from cupric oxide (CuO) revealed a decrease in the emitted signal by \SI{\sim10}{\percent} over a time interval of \SI{340}{\micro\second}, indicating FEL-induced effects that require monitoring when conducting high-repetition-rate experiments. These results establish the JUNGFRAU-iLGAD as a promising detector to harvest the full potential of the hRIXS spectrometer, and validate its suitability for soft X-ray applications.
\end{abstract}


\section{Introduction}
\label{sec01}

Spectroscopic techniques using soft X-rays are powerful probes for studies of the microscopic electronic and magnetic properties of materials. Their energy range matches the \emph{K}-edge of common light organic elements: carbon, nitrogen, and oxygen. In the same energy range we also find the \emph{L}-edges of transition metals that are often present in complex solids, including magnets, functional semiconductors and high-temperature superconductors~\cite{Fink2013}. \par

Resonant Inelastic X-ray Scattering (RIXS) stands out as a photon-in photon-out technique capable of probing charge, spin, orbital, and lattice excitations in matter~\cite{Ament2011}. By tuning the incident X-ray energy to match the absorption edge of a selected element, the inelastic scattering cross-section of the system is greatly enhanced and a unique way of accessing information about its intrinsic excitations is provided by measuring the change in energy, momentum, and polarization of the scattered photon. A particularity of RIXS is that it has an inherently low cross-section, requiring a very high incident photon flux and efficient photon detection to achieve meaningful statistics in a reasonable time. The recent emergence of extremely bright, high-repetition-rate X-ray free-electron lasers (XFELs), provided RIXS with a renewed momentum, opening up possibilities for time-resolved studies that were previously unattainable. \par

The European XFEL is an international user research facility located in Schenefeld, in the Hamburg area, that started operation in 2017. It currently features three self-amplified spontaneous emission (SASE) undulator sources, delivering up to 2700 ultra-short, spatially coherent X-ray pulses of tunable energy at a repetition rate up to \SI{4.5}{\mega\hertz}, enveloped in pulse trains of \SI{600}{\micro\second} length, that are produced at a rate of \SI{10}{\hertz}~\cite{Decking2020}. With photon energies tunable from approximately 260~\si{\electronvolt} to 25~\si{\kilo\electronvolt}, it covers a wide range of scientific possibilities. \par

Of the seven experimental stations of the European XFEL, three are dedicated to soft X-ray science~\cite{Gerasimova2022}. Among these is the Spectroscopy and Coherent Scattering (SCS) instrument, which enables time-resolved studies of electronic, magnetic and structural properties of complex materials, molecules, and nanostructures. Its scientific goals include the investigation of ultra-fast magnetization processes, the observation of chemical reactions in liquids, creation of novel transient states and the exploration of nonlinear X-ray spectroscopy techniques~\cite{SCSwebsite}. An essential instrument enabling these investigations is the Heisenberg RIXS (hRIXS) spectrometer, proposed by a user consortium and designed to be the first high-resolution spectrometer at an FEL. The hRIXS is capable of performing momentum-resolved RIXS measurements in the time domain, with time and energy resolution approaching the Heisenberg limit imposed by the uncertainty relations~\cite{Schlappa2025}. It can be coupled with two sample environment chambers: a chemistry setup (CHEM) for liquid-jet samples and an X-ray resonant diffraction (XRD) chamber for solid samples, both equipped with state-of-the-art instrumentation and compatible with optical laser pumping. \par

Despite these advanced capabilities, finding a detector capable of fully exploiting the potential of the hRIXS spectrometer remains a challenge. Particularly, this detector would have to offer high spatial resolution along the dispersive axis of the grating spectrometer, as this is one parameter directly determining the energy resolution achievable in RIXS measurements. At the same time, it must maintain low noise to reliably detect single-photon hits in the soft X-ray range, particularly between \SI{500}{\electronvolt} and \SI{1}{\kilo\electronvolt}. Additionally, to exploit the exceptional time resolution provided by European XFEL pulses, the detector needs to be capable of handling high frame rates. Finally, a large sensitive area is also important to maximize the detection solid angle and the photon collection statistics per shot. In the past, a commercial charge-coupled device (CCD)~\cite{PIMT3website} was used, while currently the hRIXS spectrometer uses the Complementary Metal-Oxide-Semiconductor (CMOS) Marana-X detector~\cite{MaranaXwebsite}. These small-pixel cameras provide good signal-to-noise ratio (SNR) and spatial resolution, but are limited in frame rate due to their slow readout time and the need for mechanical or rolling shutters. While the Marana-X detector meets the minimum requirement of resolving individual pulse trains, enabling improved time resolution through train-to-train jitter correction, the possibility to resolve individual pulses or groups of pulses within a pulse train is out of its capability. Achieving this would greatly benefit data collection and analysis, by improved options for sorting/filtering of data and e.g. allowing the use of alternating pumped and unpumped pulses for optimum data normalization in pump-probe experiments. While both CMOS and CCD technologies have undergone development in the direction of higher frame rate capabilities, such as the PERCIVAL CMOS detector~\cite{Correa2023}, the pnCCD~\cite{Struder2010, Kuster2021}, and the FastCCD~\cite{Januschek2016, Klackova2019}, they are still far from the \si{\kilo\hertz} range required to achieve intra-train resolution. A noteworthy development is the VeryFastCCD~\cite{Goldschmidt2023}, a follow-on to the FastCCD, that employs the same sensor fabrication and processing methods. Prototypes of this new generation have demonstrated a frame rate of \SI{5}{\kilo\hertz} for \(512\!\times\!512\) pixel modules, albeit with larger pixels of \SI{48}{\micro\meter}, which are relatively large for high-resolution spectroscopy applications. \par

In order to comply with the unique timing structure of the European XFEL pulse delivery, several 2D hybrid-pixel imaging detectors with wide dynamic range and frame rate capability up to \SI{4.5}{\mega\hertz} have been specifically developed by three dedicated detector consortia~\cite{Graafsma2009}. Two of these were designed for hard X-rays: the Adaptive Gain Integrating Pixel Detector (AGIPD)~\cite{Allahgholi2019}, and the Large Pixel Detector (LPD)~\cite{Veale2017}. The third one is the DEPFET Sensor with Signal Compression (DSSC)~\cite{Porro2021} detector, which despite featuring single photon sensitivity in the soft X-ray regime, does not possess the required spatial resolution due to its large hexagonal pixels of \SI{\sim200}{\micro\meter}. \par

In recent years, there have been attempts to increase the SNR of hybrid detectors by combining established readout architectures designed for hard X-rays with sensor elements capable of providing signal amplification by charge multiplication. This resulted in the employment of Low Gain Avalanche Diode (LGAD) sensors~\cite{Andraë2019}, originally developed over a decade ago for high-energy physics~\cite{Pellegrini2014}. In particular, a collaboration between the Paul Scherrer Institute (PSI) and Fondazione Bruno Kessler (FBK) has materialized into the first prototypes of soft X-ray-sensitive inverse Low Gain Avalanche Diode (iLGAD) sensors~\cite{Zhang2022} compatible with the readout chips of the widely adopted charge-integrating detectors developed by PSI, namely the JUNGFRAU~\cite{Mozzanica2018}, GOTTHARD-II~\cite{Zhang2021} and MÖNCH~\cite{Dinapoli2014}. \par

A JUNGFRAU detector equipped with an iLGAD sensor was recently characterized and tested at the hRIXS spectrometer of the European XFEL. In this work, we present a detailed characterization of the detector, focusing on noise performance, gain, SNR, and signal uniformity across memory cells. We also show the first RIXS spectra acquired at multi-kHz rate in combination with pump-probe optical laser excitation. Using reference solid samples, we highlight the achieved spatial resolution and resolving power, comparing them to the previously used commercial cameras. Furthermore, we report how the intense FEL pulses can influence the detected signal from the sample, offering an important insight to optimize experimental settings and RIXS data acquisition, as well as identifying potential limitations in such measurements. \par

\section{Detector properties}
\label{sec02}

A detector that fully exploits the capabilities of the hRIXS spectrometer at the European XFEL must possess certain characteristics. First and foremost, to ensure the best possible energy resolution that the experimental apparatus is capable of, it should provide an estimate of each photon hit position with an accuracy of \SI{5}{\micro\meter} or less in the dispersive direction. It should also offer single photon sensitivity for energies below \SI{1}{\kilo\electronvolt}, down to \SI{\sim400}{\electronvolt}. Finally, to best exploit the light delivery structure of the FEL, it should be capable of burst frame rate of \SI{1.1}{\mega\hertz}; lower frame rates in the order of \SI{100}{\kilo\hertz} are also useful, as they still offer the so far unprecedented possibility of intra-train measurements. In this section the characterization of the properties of the JUNGFRAU detector equipped with an iLGAD sensor is presented, quantifying its compliance with the aforementioned requirements. \par

\subsection{JUNGFRAU detector with iLGAD sensors}

The JUNGFRAU detector system is a hybrid-pixel detector developed by PSI for the Swiss Free-Electron Laser (SwissFEL) and currently used in many synchrotron and XFEL facilities worldwide. At the European XFEL, several modules are deployed and routinely used across all hard X-ray instruments, in experiments from protein crystallography~\cite{Tolstikova2019} to spectroscopy~\cite{Preston2020}. They extend the scientific possibilities of the instruments by providing an easily maneuverable and smaller-pixel alternative to the large MHz-rate detectors specifically developed for the European XFEL. Here, the fundamental characteristics of the JUNGFRAU are summarized, while a detailed description can be found in~\citeasnoun{Mozzanica2014}, and a review of its performance specifically for European XFEL applications can be found in~\citeasnoun{Sikorski2023}. \par

A JUNGFRAU ASIC consists of \(256\!\times\!256\) square pixels with a pitch of \SI{75}{\micro\metre}, making for approximately \(2\!\times\!2\)~\si{cm\squared}. In the standard design, \(2\!\times\!4\) chips are tiled to form a \(4\!\times\!8\)~\si{cm\squared} module of around 0.5 megapixels, bump-bonded to a planar p-on-n silicon sensor \SI{320}{\micro\meter} or \SI{450}{\micro\meter} thick. Each pixel features a charge integrating preamplifier with three possible amplification factors, referred to as \emph{high gain} (G0), \emph{medium gain} (G1), and \emph{low gain} (G2). The preamplifier gain is automatically selected during acquisition by a dynamic switching logic based on a pixel-wise threshold comparator. The output of the preamplifier is then sampled by a Correlated Double Sampling stage (CDS) to reduce the signal noise. The CDS stage itself provides a further amplification factor. Moreover, the G0 gain can be statically increased by disconnecting part of its feedback capacitance. By doing so, an additional gain configuration called \emph{high gain0} (HG0) is selected, which is particularly useful to enhance the SNR ratio in applications with low photon fluxes or low photon energy. The charge is stored in analog memory cells before being digitized and read out together with the corresponding gain-bit value used for coding the different amplification factors for every pixel in-between pulse trains. This pixel design allows for a dynamic range that extends from single photon sensitivity up to $\sim 10^{4}$ \SI{12}{\kilo\electronvolt} photons, a capability that is not exploited in RIXS measurements due the extremely low photon yield inherent to the technique. \par

The original chip version (v1.0) offers a burst mode acquisition with 16 memory cells, and it is therefore capable of recording up to 16 frames within a single European XFEL pulse train, reaching a burst acquisition rate up to \SI{\sim200}{\kilo\hertz}. Operated in HG0, it achieves an equivalent noise charge (ENC) noise of \SI{\sim50}{e^-} r.m.s. at a \SI{5}{\micro\second} integration time, resolving singe photons down to \SI{\sim1.5}{\kilo\electronvolt} with an SNR of \SI{\sim8}{}~\cite{Mozzanica2016}. A more recent chip version (v1.1) improves noise performance by reducing the capacitance of the preamplifier, the feedback capacitance in high gain, and by adding a filtering resistor before the CDS stage. With these improvements, noise is lowered by \SI{\sim35}{\percent}, down to \SI{34}{e^-} r.m.s. at a \SI{5}{\micro\second} integration time~\cite{Hinger2022}. To achieve so, however, a number of memory cells were converted for filtering, reducing their effective number to only 4. For the hRIXS spectrometer, maximizing intra-train resolution is prioritized, making the v1.0 chip version preferable, since it can provide a higher number of frames per train. Moreover, with only 4 gates, v1.1 would require longer integration times for each gate to cover the typical \SI{\sim340}{\micro\second} FEL train window of the SCS instrument, which would in turn increase noise. A new chip version (v1.2), currently at an advanced development stage, is able to achieve even lower noise (\SI{\sim31}{e^-} r.m.s.) while reimplementing the 16 memory cells, by using a fixed filter capacitance~\cite{Kedych2025}. This chip is the ideal candidate for future JUNGFRAU-iLGAD prototypes. \par

In order to make the JUNGFRAU a suitable option for soft X-ray applications, the loss of quantum efficiency for low-energy photons must initially be addressed. The standard sensor of the JUNGFRAU has a quantum efficiency of \SI{60}{\percent} for \SI{900}{\electronvolt} photons, which rapidly decreases to only \SI{10}{\percent} at \SI{500}{\electronvolt}~\cite{Zhang2022}. This is due to the fact that for such low energies, photons mostly interact in the first few micrometers of the sensor, where an entrance window is located: an insensitive layer combined with a highly doped region, typically ranging in thickness from a few hundred nanometers to a couple micrometers~\cite{Carulla2023}. The entrance window serves the purpose of protecting the silicon sensor, and providing a good high voltage contact structure, generating a stable and uniform electric field across the sensor. However, this either causes photons to be absorbed in the insensitive layer of the entrance window, or the generated charges to recombine in the highly doped region or at the silicon surface. These photons either do not produce a signal in the detector or produce a very small signal due to partial charge collection, leading to a decrease in quantum efficiency. To overcome this, so-called thin entrance windows have been investigated and developed, using different fabrication processes to push the thickness of the entrance windows to the limit of what is technologically achievable. Already with the earlier prototypes of this optimized processes, a quantum efficiency of \SI{90}{\percent} for \SI{900}{\electronvolt} and \SI{65}{\percent} for \SI{500}{\electronvolt} photons has been measured~\cite{Hinger2022}, which constitutes a significant improvement. With more recent prototypes, featuring a thin entrance window with passivated surface, quantum efficiency values reach \SI{82}{\percent}-\SI{93}{\percent} for \SI{500}{\electronvolt} and \SI{62}{\percent}-\SI{80}{\percent} for \SI{250}{\electronvolt} photons, depending on different shallowness of the passivation layer~\cite{Carulla2024}. \par

The second challenge in advancing hybrid-pixel detectors towards soft X-ray applications is the fact that the primary signal generated in the sensor by low-energy photons is not sufficient to overcome the electronic noise. As an example, considering the mean electron-hole pair creation energy of \SI{3.6}{\electronvolt} in silicon, a \SI{500}{\electronvolt} photon produces, on average, only about \SI{\sim140}{} initial electron-hole pairs. This is less than a factor of 3 above the ENC noise of \SI{\sim50}{e^-} in JUNGFRAU 1.0 chips. A strategy to overcome this is to amplify the signal in the sensor itself through a charge avalanche multiplication process. In their original design~\cite{Pellegrini2014}, LGADs achieve this by implanting a highly doped p-type region below the n-type readout electrodes, referred to as multiplication or gain layer. When the sensor is biased, a local electric field of several hundred \SI{}{\kilo\volt/\cm} is established in this region, high enough for charge multiplication through impact ionization to occur, allowing for a multiplication factor of up to 10. \par

However, the standard LGAD design is not suited for application in photon science. The primary reason for this is that by implementing the multiplication layer below the readout electrodes, the multiplication layer is itself segmented~\cite{Andraë2019}. This results in a limited fill factor (ratio between multiplication layer surface and total sensor surface), since primary charge carriers generated from photon absorptions in regions not aligned with the electrodes will not undergo multiplication. As a consequence, gain is not uniform across the sensor, which compromises energy resolution and makes the use of interpolation methods through charge sharing unfeasible. With an alternative sensor design, the inverse LGAD (iLGAD)~\cite{Pellegrini2016}, this issue is resolved by implementing the multiplication layer on the rear side, immediately after the entrance window, covering all readout electrodes within the entire pixel area. In this configuration, photons are ideally absorbed after traversing the entire multiplication layer, generating the initial electron-holes pairs. The electrons then drift back toward the multiplication layer where they trigger the charge multiplication process that produces charge carriers. The generated holes, together with those produced by X-rays directly, drift to the collecting electrodes to generate most of the detection signal. While this results in a longer detection time due to the lower mobility of holes compared to electrons, this drawback is not critical for photon science experiments, making the iLGAD a promising option. Another consequence of this design is that a fraction of photons may be absorbed before fully traversing the multiplication layer, leading to hole-initiated avalanches that result in a lower gain due to their smaller impact ionization coefficient compared to electrons. Because lower-energy photons are more likely to be absorbed within the multiplication layer, a dependence between the multiplication factor and photon energy is introduced, a limitation that can be mitigated by reducing the thickness of the multiplication layer. \par

Due to the high flexibility that hybrid-pixel detectors provide, it is relatively straightforward to combine an iLGAD sensor with a JUNGFRAU chip. However, two operational requirements should be fulfilled. The first one is the need for higher bias voltage (around \SI{300}{\volt}) to optimize the electric field magnitude for avalanche multiplication in the gain layer. The second requirement is cooling the sensor, as SNR has been measured to improve exponentially with decreasing temperatures~\cite{Hinger2024}. This improvement is due to two factors: first, the multiplication factor increases as the sensor is cooled; second, leakage current, the thermally generated current that flows through the sensor even in the absence of incident radiation, decreases significantly at lower temperatures. Because charges arising from leakage current are also amplified in the gain layer, it is a key component to noise in iLGADs, making its suppression essential to improve SNR. \par

\begin{figure}[!htb] %
\begin{center}
\includegraphics[width=0.5\textwidth]{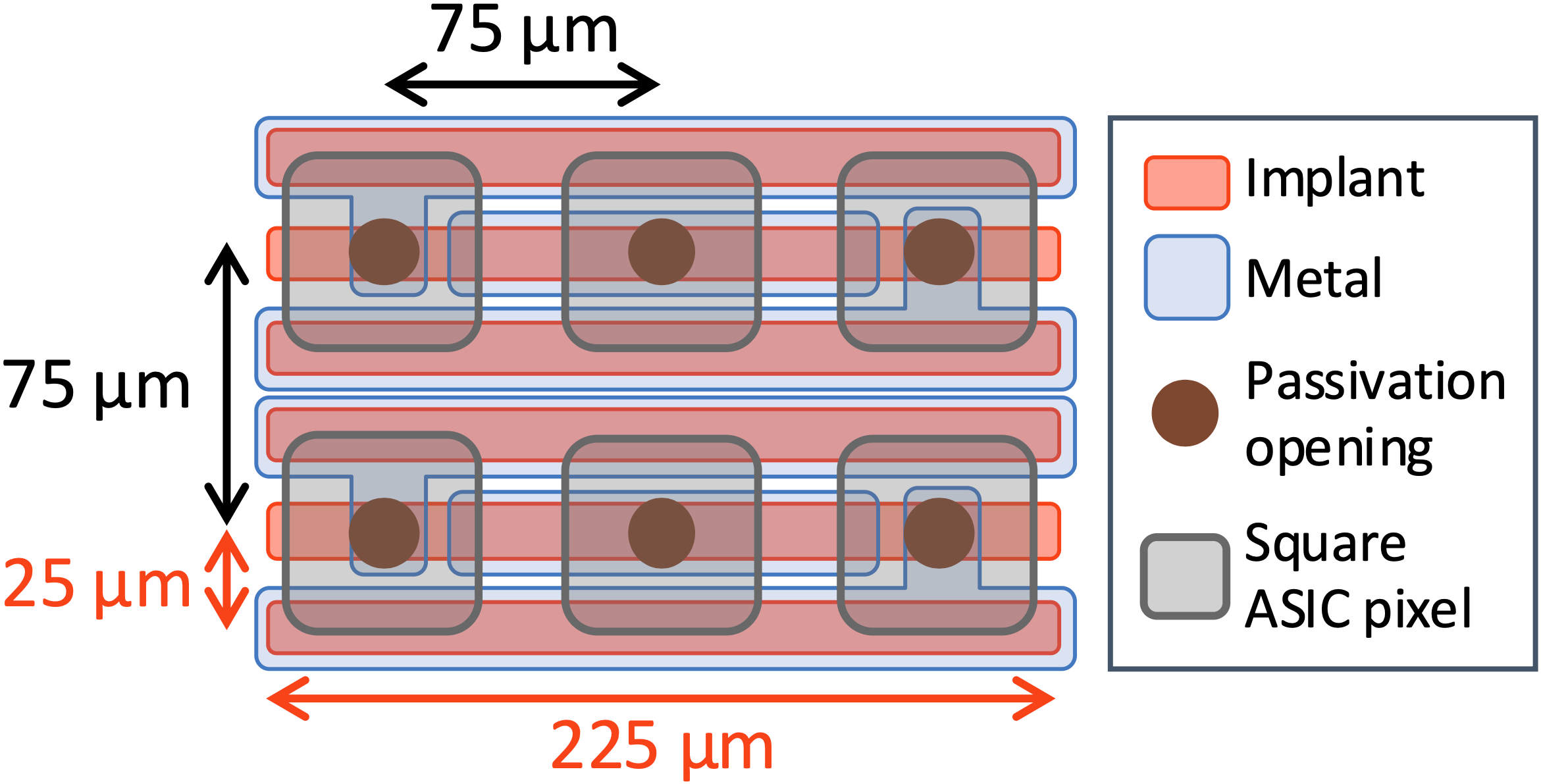}
\end{center}
\caption{Strixel sensor geometry with the elongated \(25\!\times\!225\)~\si{\micro\metre\squared} pixel design and corresponding mapping between the multiplication layer implants and the original JUNGFRAU ASIC pixel.}
\label{fig:sec02_strixel}
\end{figure}

Additionally, the sensor can be segmented into narrow rectangular pixels of \(25\!\times\!225\)~\si{\micro\metre\squared} (or narrower), while preserving the original pixel layout of \(75\!\times\!75\)~\si{\micro\metre\squared} at the ASICs level (Fig~\ref{fig:sec02_strixel}). This is achieved by reducing the pixel size by a factor of 3 (or more) in one dimension, while extending it by the same factor in the other dimension (forming so-called ``strixels"), thus preserving the total pixel area and the number of bump-bonds. This segmentation requires a modified post-processing procedure which maps each ASIC pixel to the corresponding position of its bond to the sensor, ensuring it is accurately represented. Through this process, the spatial resolution is increased along the energy dispersion axis for spectroscopic experiments, and can be further enhanced by leveraging charge sharing across pixels through position interpolation methods, as will be discussed in section~\ref{subsec:charge_sharing}. \par

\subsection{Detector characterization}

\begin{figure}[!htb] %
\begin{center}
\includegraphics[width=0.5\textwidth]{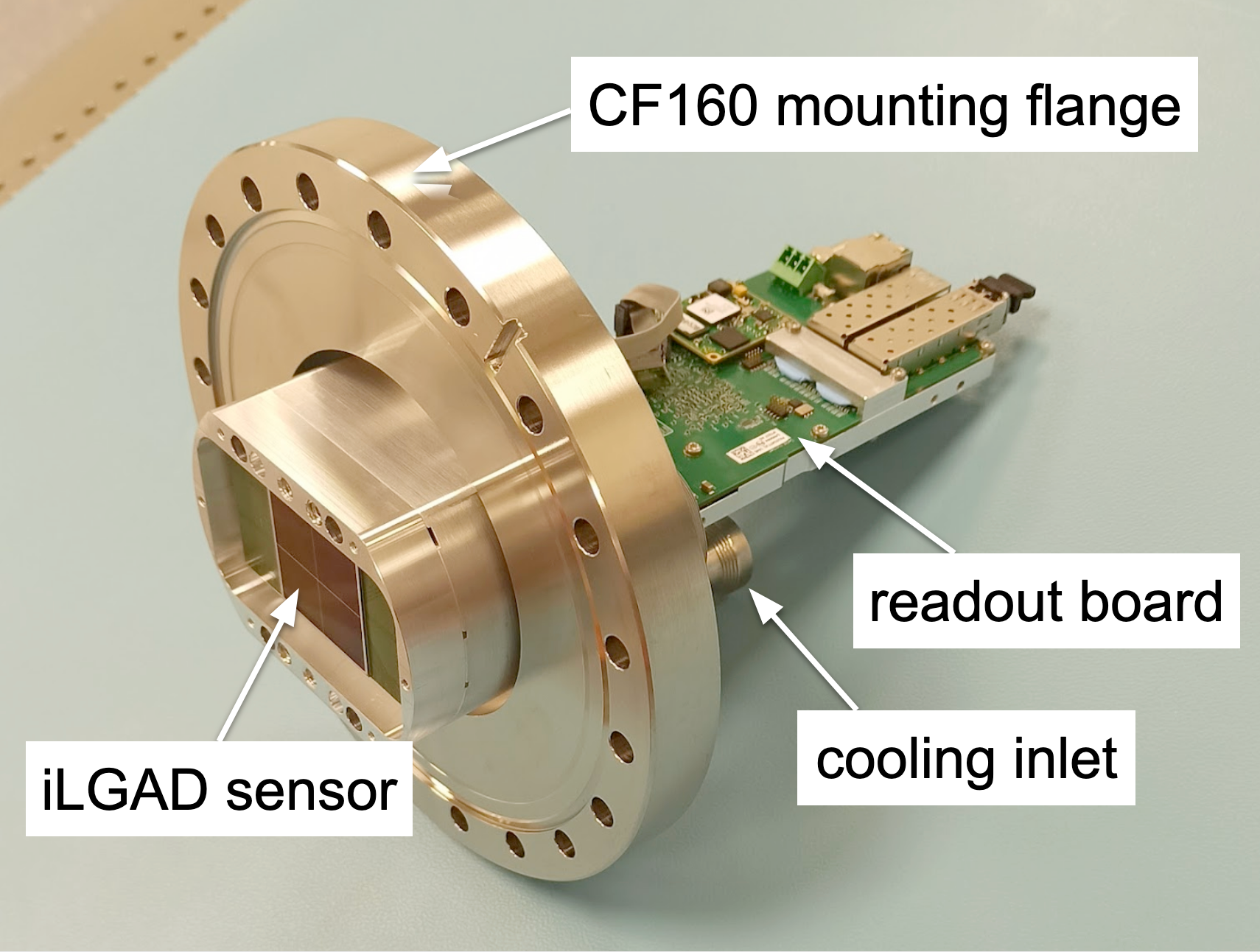}
\end{center}
\caption{JUNGFRAU-iLGAD detector used in this work, connected to a 160~CF flange for vacuum integration. Key components are the front-end module with the iLGAD sensor and ASICs, the readout board, and coolant inlet and outlet (not visible).}
\label{fig:sec02_JF_picture}
\end{figure}

The prototype used in this work was a JUNGFRAU with ASIC v1.0, bump-bonded to an iLGAD sensor with total thickness of \SI{275}{\micro\meter}. The sensor features a thin entrance window around \SI{\sim80}{\nano\meter} thick, followed by a standard gain layer around \SI{\sim670}{\nano\meter} thick~\cite{Liguori2023}. The detector was composed of a \(2\!\times\!2\) ASIC array, and rectangular pixels with a size of \(25\!\times\!225\)~\si{\micro\metre\squared}, making for a total sensitive area of \(\sim4\!\times\!4\)~\si{cm\squared}. A photograph is shown in Fig.~\ref{fig:sec02_JF_picture}, including the assembly with a ConFlat (CF) flange for vacuum integration. \par

Prior to its installation in the hRIXS spectrometer, a measurement campaign for detector calibration and characterization was carried out. For this purpose, an X-ray generator was used, consisting of an electron gun with tunable energy and intensity, followed by a beam blanker capable of generating a continuous X-ray beam or X-ray pulses with a time structure similar to that of the European XFEL, by directing electrons onto a multi-target wheel~\cite{Kuster2025}. An additional filter wheel with foils of different materials and thicknesses can be used to attenuate the fluorescence emitted from the target, allowing the energy spectrum to be shaped or the X-rays to be monochromatized before reaching the detector. \par

Throughout these measurements, the sensor was cooled using a chiller set to a bath temperature of \SI{-25}{\degreeCelsius}, with a continuous flow of silicone oil as coolant, and biased using a high voltage board mounted on an MPOD crate~\cite{Wienerwebsite}. The detector was operated in \emph{high gain0} (HG0), since this gain configuration optimizes SNR for low energy photons. Different bias voltages were applied to assess the impact on detector performance: \SI{200}{\volt}, \SI{250}{\volt}, and \SI{300}{\volt}. \par

\subsubsection{Noise}

The noise characterization of a detector is crucial for benchmarking its performance, understanding its limitations, and tracking its stability over time. The overall detector noise arises from multiple sources, including the electronic circuitry, the sensor itself, and the statistical nature of charge carrier generation. In iLGAD sensors, leakage current plays a critical role in the overall noise contribution, as thermally generated charges are also amplified by the gain layer. When evaluating detector noise, we are typically concerned with the total noise ($\sigma$), which can be quantified for each pixel by analyzing fluctuations in its offset, i.e, the output signal measured in the absence of incident photons. Pixel noise is therefore commonly defined as the standard deviation of this offset. \par

\begin{figure}[!htb] %
\begin{center}
\includegraphics[width=0.9\textwidth]{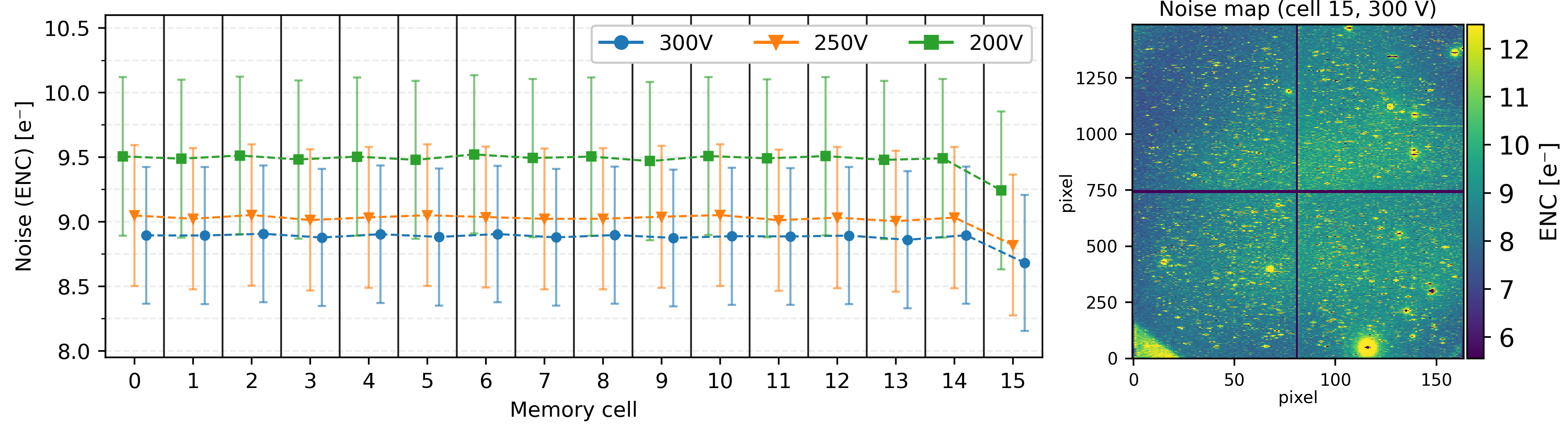}
\end{center}
\caption{Left: Average noise per memory cell for different sensor bias voltages, measured from dark frames with \SI{5}{\micro\second} integration time. Cell 15 exhibits lower noise due to larger physical capacitance. Right: Pixel-wise noise map for memory cell 15 at \SI{300}{\volt}.}
\label{fig:sec02_noise}
\end{figure}

The noise performance of the JUNGFRAU-iLGAD used in this work is presented in Fig.~\ref{fig:sec02_noise}-left, measured from 1000 dark frames taken with an integration time of \SI{5}{\micro\second}, for different bias voltages. Conversion from arbitrary detector units (ADU) to equivalent noise charge (ENC) is done using the gain constant ($g$) calculated from flat-field measurements (as explained in section~\ref{sec02:gain}) and the mean electron-hole pair creation energy in silicon: \par

\begin{equation}
\mathrm{ENC}~[\si{e^-}]= \frac{\sigma~[\si{ADU}]}{g~[\si{ADU}/\si{eV}] \times 3.6~[\si{eV}/\si{e^-}]} 
\label{eq:ENC}
\end{equation}

The lowest noise is achieved for a bias voltage of \SI{300}{\volt}, with an average ENC value of $8.9\pm1.1~\si{e^-}$ across all pixels (excluding bad pixels as defined in section \ref{sec02:bad_pixels}) and memory cells, with good uniformity between pixels across the sensor plane, as exemplified in the noise map of Fig.~\ref{fig:sec02_noise}-right for memory cell 15 at \SI{300}{\volt}. The verified trend indicates that the multiplication factor of the iLGAD sensor increases with the bias voltage, and that the resulting increase in photon signal outweighs the additional noise introduced by leakage current amplification and fluctuations in the avalanche multiplication process of said leakage current. The 16 memory cells show a uniform noise level, with the exception of memory cell 15, which has \SI{\sim 5}{\percent} lower noise. This corresponds to the default memory cell connected to the readout circuitry when the JUNGFRAU is used in single cell mode, having a ~20\% larger capacitor and showing the best performance. \par

\subsubsection{Gain and signal-to-noise ratio}
\label{sec02:gain}

In addition to the conventional electronic amplification of the pixel signal, the multiplication layer in the iLGAD sensor also contributes to the overall detector gain. Dedicated measurements using modified iLGAD sensors (in which a portion of the sensor is intentionally left without a multiplication layer) allow each of the contributions to the total gain to be disentangled. Such studies, conducted on a similar iLGAD sensor as the one used in this work, have reported a multiplication factor ranging from approximately 11 to 15 at \SI{-22}{\degreeCelsius}~\cite{Hinger2024}. Here, we focus only on the overall detector gain, as this is the relevant factor for determining the SNR. \par

\begin{figure}[!htb] %
\begin{center}
\includegraphics[width=0.9\textwidth]{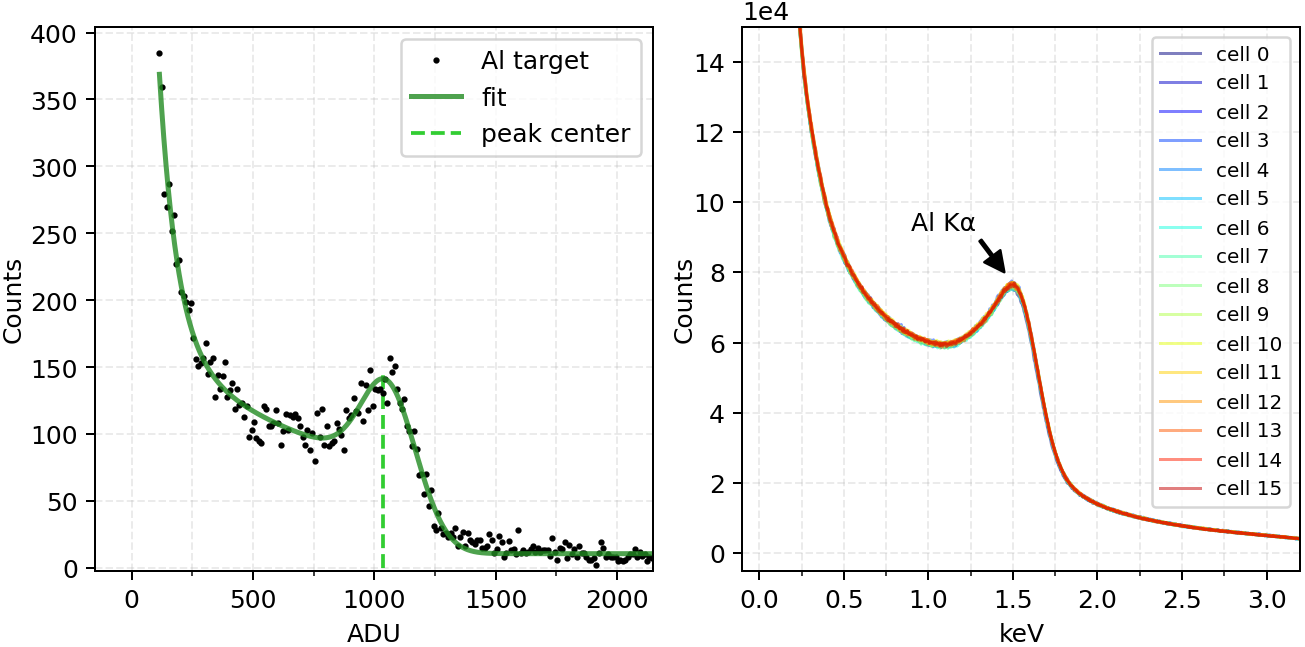}
\end{center}
\caption{Left: Energy spectrum of a single pixel under uniform illumination of \SI{1.49}{\kilo\electronvolt} Al K-shell fluorescence photons, at \SI{300}{\volt} sensor bias. Right: Summed energy spectra from all pixels, grouped by memory cell, after gain calibration, indicating a uniform energy response.}
\label{fig:sec02_pulxar}
\end{figure}

To characterize the detector gain, the sensor is uniformly illuminated with photons of known energy (flat-field), at a sufficiently low flux to maximize the likelihood that each pixel integrates no more than a single photon per integration gate. The electron gun was set to produce and accelerate electrons with a \SI{10}{\kilo\volt} electric field, and an aluminum target was used to generate \SI{1.49}{\kilo\electronvolt} K-shell fluorescence photons. The energy spectrum collected by a single pixel after pedestal subtraction is shown in Fig.~\ref{fig:sec02_pulxar}-left. The spectrum is trimmed on the low-energy side to emphasize the photon peak, which reveals that the pixel response to single-photon illumination does not follow a simple Gaussian curve. Instead, a more complex shape with a left-side tail is observed, attributed to two factors. First, bremsstrahlung photons generated by electrons striking the aluminum target also reach the detector, with a broad energy distribution extending down to zero. Second, due to the small pixel pitch, there is a prominent charge sharing effect that contributes to the low-energy population of the spectrum, as discussed in detail in section~\ref{subsec:charge_sharing}. The pixel energy response is best described by a combination of an exponential decay function and a charge-sharing function~\cite{Cartier2016}. The position of the local maximum (peak) of the fitted curve is used as a reference to determine the conversion ratio between the measured ADU value and the corresponding photon energy in \si{keV} for that specific pixel (gain constant). By applying this procedure across all pixels and memory cells, a gain map is generated. This map is then applied to pedestal-corrected data to compensate for pixel-to-pixel gain variations and to convert their outputs to energy units. After this correction, a uniform energy response is achieved across all memory cells, as illustrated in Fig.~\ref{fig:sec02_pulxar}-right. The figure shows the summed energy spectra from all pixels for each memory cell under illumination from the aluminum target. Following gain calibration, the spectral lines from all memory cells are superimposed and virtually indistinguishable, demonstrating the uniform response across the 16 memory cells. \par

\begin{figure}[!htb] %
\begin{center}
\includegraphics[width=0.9\textwidth]{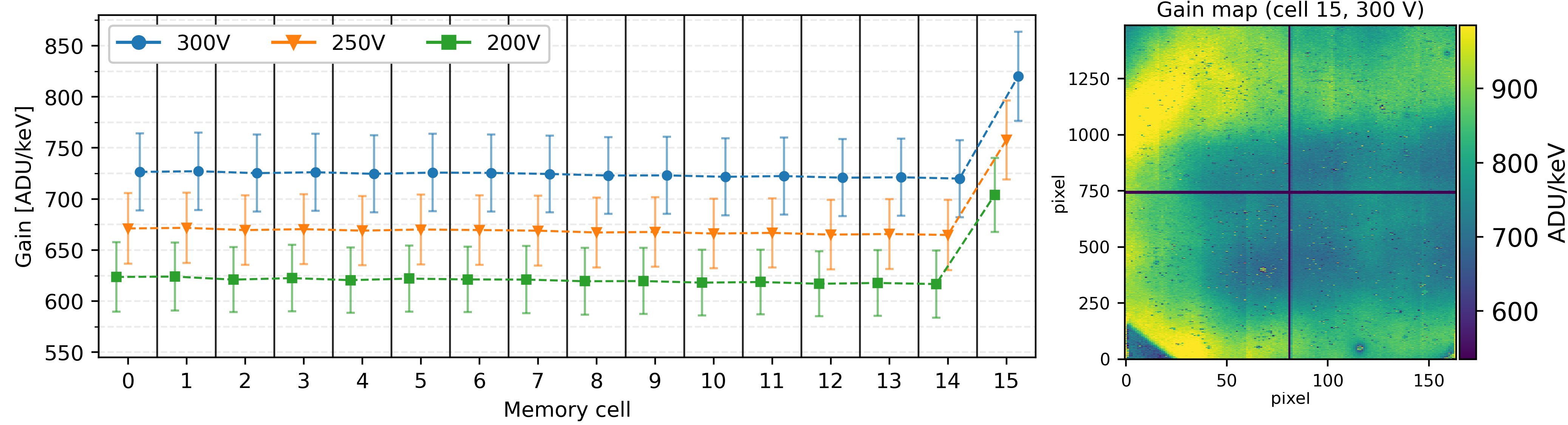}
\end{center}
\caption{Left: Average gain constant per memory cell at \SI{200}{\volt}, \SI{250}{\volt}, and \SI{300}{\volt} bias voltage. Right: Pixel-wise gain map for memory cell 15 at \SI{300}{\volt}.}
\label{fig:sec02_gain}
\end{figure}

The calculated gain constants are highly uniform across memory cells, as shown in Fig.~\ref{fig:sec02_gain}-left, which displays the average gain constant calculated across all pixels for each memory cell. Similar to the behavior observed in the noise characterization, memory cell 15 appears as an outlier, exhibiting a higher gain constant, due to its 20\% larger capacitance. It is also observed that the gain increases with bias voltage, which is attributed to the fact that primary charges accelerated by stronger electric fields acquire more energy and generate a larger number of secondary charges through avalanche multiplication in the gain layer, consequently increasing the multiplication factor. As for spatial uniformity of gain across individual pixels, illustrated in Fig.\ref{fig:sec02_gain}-right for memory cell 15, moderate fluctuations across the sensor area are observed, as is not uncommon in hybrid-pixel detectors. In this case, the variations are likely to arise from both electronic differences between pixels and non-uniformities in the gain layer across the sensor plane. \par

\begin{figure}[!htb] %
\begin{center}
\includegraphics[width=0.8\textwidth]{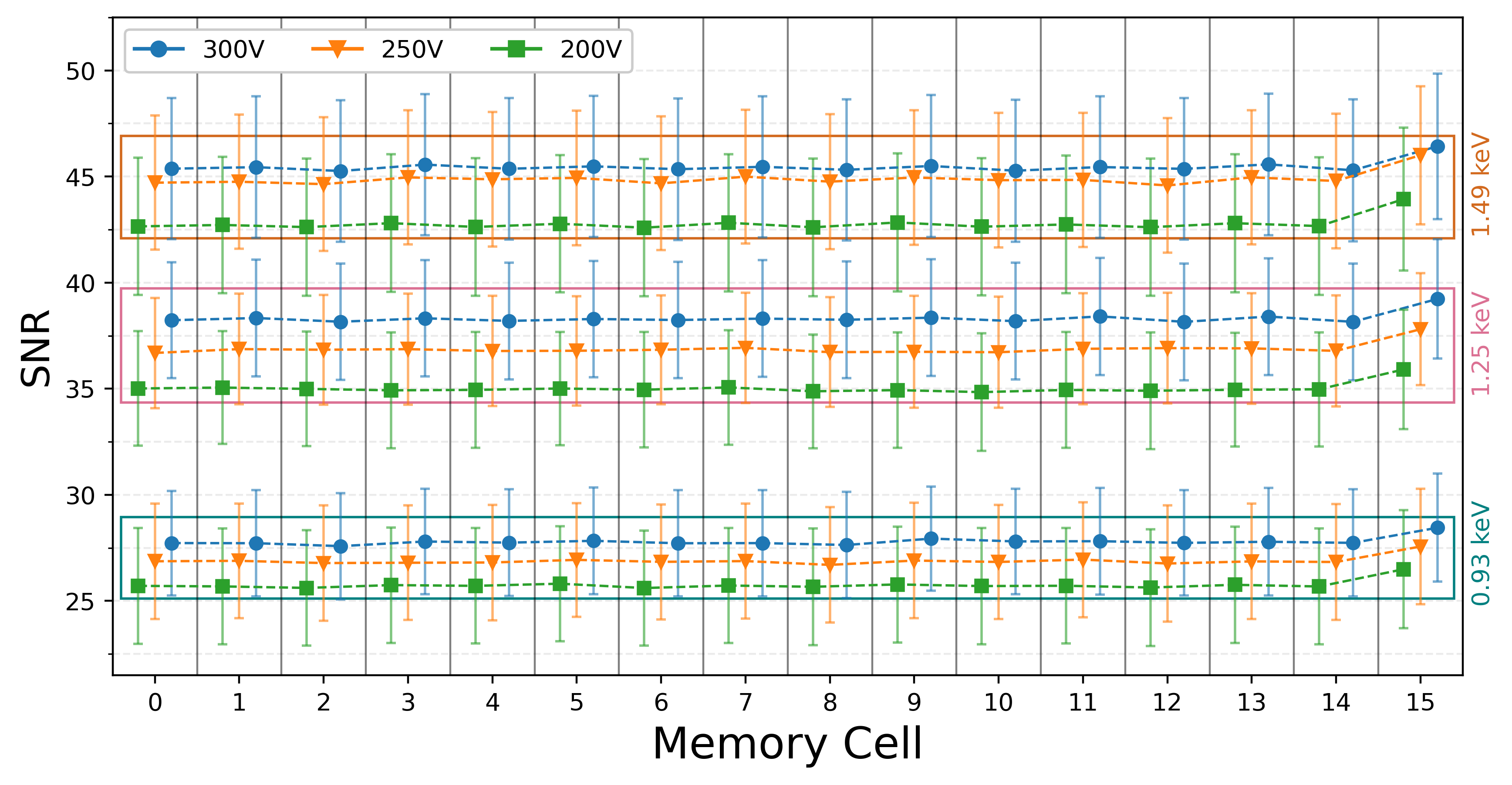}
\end{center}
\caption{SNR as a function of memory cell for different photon energies and sensor bias voltages at \SI{5}{\micro\second} integration time.}
\label{fig:sec02_SNR}
\end{figure}

The SNR was assessed using dark frames and flat-field measurements acquired at three different sensor bias voltages: \SI{200}{\volt}, \SI{250}{\volt}, and \SI{300}{\volt}. To study the variation of SNR with photon energy, flat-field exposures were performed using aluminum, magnesium, and copper targets, which emit fluorescence photons at \SI{1.49}{\kilo\electronvolt}, \SI{1.25}{\kilo\electronvolt} and \SI{0.93}{\kilo\electronvolt}, respectively. Integration time was kept at \SI{5}{\micro\second}. For each configuration, the SNR per pixel was calculated as: \par

\begin{equation}
\text{SNR} = \frac{\mu_{\text{flat-field}} - \mu_{\text{dark}}}{\sigma}
\end{equation}

where $\mu_{\text{flat-field}}$ is the position of the local maximum of the fitted peak in the flat-field measurement, $\mu_{\text{dark}}$ is the offset of the corresponding dark frame, and $\sigma$ the pixel noise. The results, shown in Fig.~\ref{fig:sec02_SNR}, reveal that SNR consistently improves with increasing bias voltage across all photon energies, as expected from the associated higher gain. Variation between memory cells remain small, typically, confirming the uniformity of detector performance in this aspect. At \SI{300}{\volt} operation, the SNR exceeds 45 for \SI{1.49}{\kilo\electronvolt}, 38 for \SI{1.25}{\kilo\electronvolt}, and 27 for \SI{0.93}{\kilo\electronvolt} photon energies. \par

\subsubsection{Bad Pixels}
\label{sec02:bad_pixels}

A precise identification of bad pixels is essential for reliable data analysis, as such pixels can output a signal that might be wrongly interpreted as a photon hit. This is especially critical when dealing with a sensor prototype still undergoing optimization processes, and even more important for photon-hungry experiments, where each detected photon carries a relatively high statistical weight. In RIXS measurements, the issue is further aggravated due to the localized distribution of photons, which are typically confined to a narrow region on the sensor plane. Consequently, a single bad pixel within this region can have a disproportionately large impact on the accuracy of the calculated spectrum. To mitigate this, more conservative thresholds for pixel rejection are applied. \par

Pixels are marked as ``bad" if their noise, offset, or gain significantly deviate ($>5$ standard deviations) from the average, or if there are not enough photon statistics for an accurate gain calibration. Because of this, both dark runs and flat-fields are used to identify bad pixels. \par

The overall fraction of bad pixels for the tested JUNGFRAU-iLGAD prototype remains below \SI{1}{\percent}, excluding the group of outlier pixels visible in the bottom-left corner of the sensor shown in Fig.~\ref{fig:sec02_gain}-right. These pixels could not be characterized to a satisfactory degree because they had minimal photon flux during flat-field measurements, a result of technical constraints in aligning the X-ray generator to obtain uniform illumination across the whole sensor surface. \par

\begin{figure}[!htb] %
\begin{center}
\includegraphics[width=.7\textwidth]{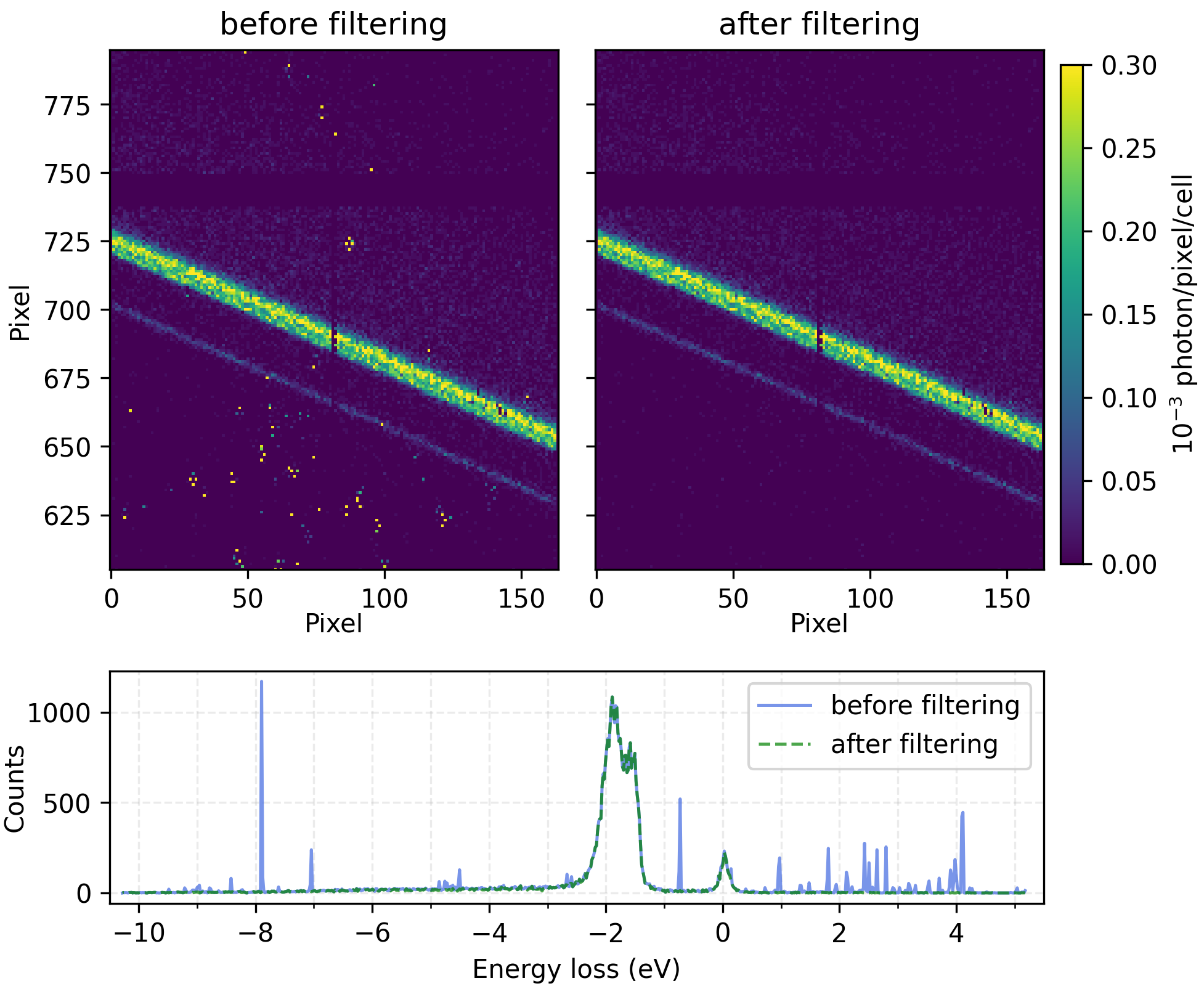}
\end{center}
\caption{Top: Comparison of the 2D images obtained after frame averaging before (left) and after (right) bad pixel filtering. Bottom: Corresponding RIXS spectrum for each case, calculated as described in section~\ref{sec03:methods}.}
\label{fig:sec02_bad_pixels}
\end{figure}

In addition to the described criteria for bad pixel identification, filtering was implemented to address observed time-dependent pixel instabilities. Specifically, a fraction of pixels exhibited nominal noise and offset values during dark characterization, but (later) showed abnormally high ADU values during RIXS measurements, which could lead to their misidentification as photon events. The identification of these unstable pixels was possible by exploiting the fact that the photon count rate on the detector is homogeneous along the non-energy dispersive direction. Hence, outlier pixels could be identified by computing median and median absolute deviation along the non-energy dispersive direction. The resulting effect on the 2D image obtained from frame averaging and on the derived RIXS spectrum is illustrated in Fig.~\ref{fig:sec02_bad_pixels}, showing that the spurious events from bad pixels can be effectively filtered out. \par

\subsubsection{Charge sharing and position interpolation}
\label{subsec:charge_sharing}

When the pitch of a segmented X-ray detector is comparable to the size of the charge cloud produced by an absorbed photon, a charge sharing effect is typically present, by which the signal from a single photon interaction is distributed across multiple pixels/strips. In photon-counting detectors, this may result in events not being recorded if the energy deposited in each pixel falls below the event detection threshold. For charge-integrating detectors, moreover, charge sharing produces a tail and a shift towards lower energies of the spectrum, arising from the contribution of pixels that integrate only a fraction of the total charge generated by the photon absorption. If not accounted for, these effects degrade spatial resolution, leading to a blurring effect in the acquired images. While small-pitch detectors are essential for applications that require high spatial resolution, this effort can ultimately be hampered by the charge sharing effect if not effectively addressed and corrected for. Additionally, charge sharing is a complex process to model \emph{a priori} or to systematically calibrate for, since it depends not only on the sensor material, thickness, temperature, pitch size, and bias voltage, but also on the incoming photon energy and flux~\cite{Becker2011, Abboud2013}. Clustering methods can be applied to compensate for this effect. These consist of identifying pixels or strips with signal corresponding to the same photon interaction and grouping them to reconstruct the full energy of the photon peak. It can be done by scanning a 1D or 2D window of predefined size across the detector pixel array and identifying candidate pixels that exceed a primary threshold relative to the noise level as potential cluster centers. For each candidate, a local window is defined around it, and the surrounding pixels are evaluated to determine if their signal exceeds a second noise threshold, indicating charge sharing. If the central pixel is the local maximum within the window, the signals of neighboring pixels above threshold are summed and assigned to the central pixel, while the values of the pixels in the cluster are set to zero. This ensures that shared charges are correctly aggregated without double-counting and provides a clean separation of individual photon events. A secondary data structure can be used to track the size and composition of each cluster, providing statistics on the determined clusters and an insight on the charge sharing prevalence and behavior. The discriminating noise thresholds can be optimized for different pixel geometries, operation parameters, and photon energy. \par

\begin{figure}[!htb] %
\begin{center}
\includegraphics[width=0.5\textwidth]{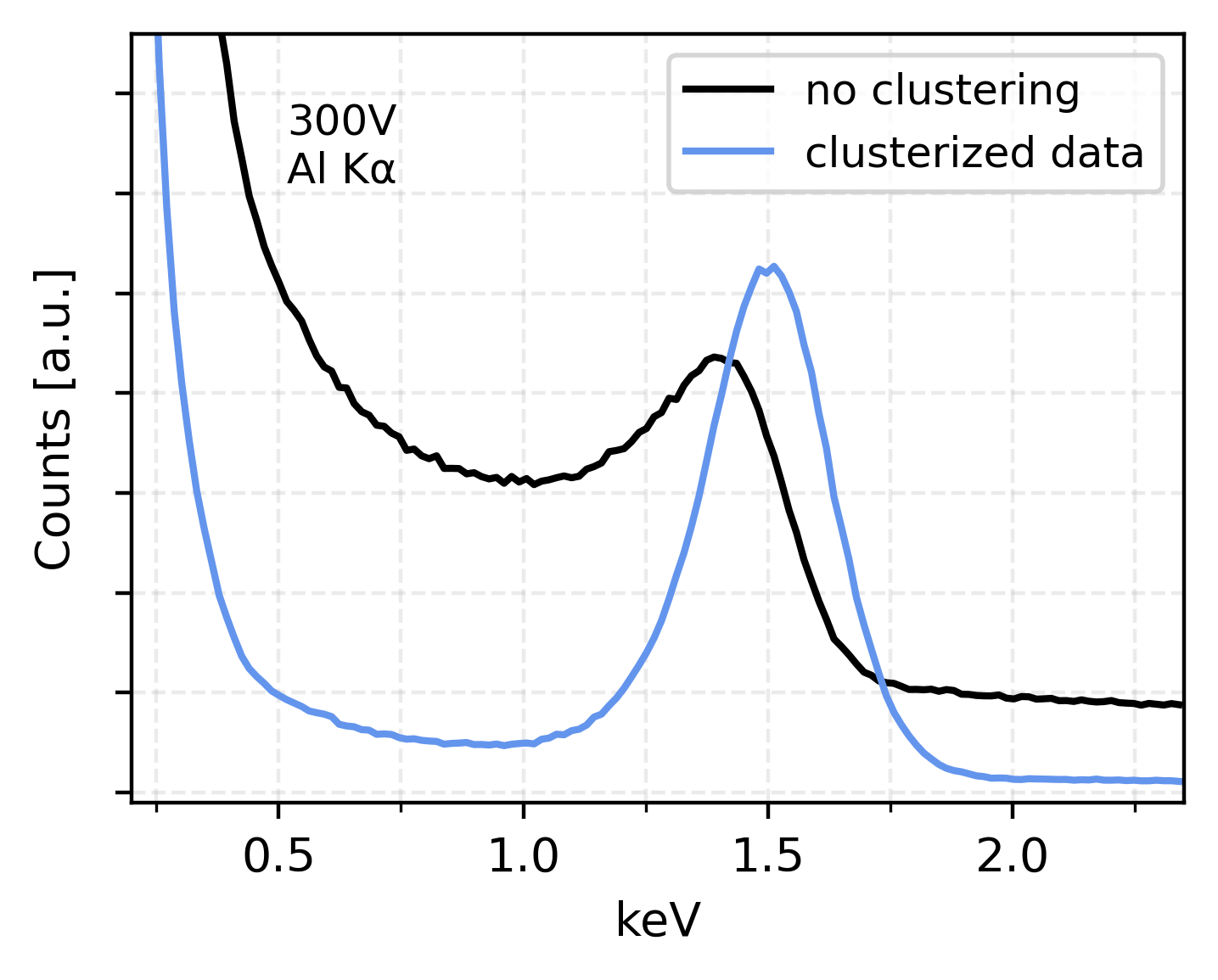}
\end{center}
\caption{Effect of charge sharing on the energy spectrum of Al K$\alpha$ X-rays at \SI{300}{V} bias, with and without clustering.}
\label{fig:sec02_charge_sharing}
\end{figure}

An example of the energy spectrum of the tested JUNGRAU-iLGAD prototype, summed across all pixels and memory cells, before and after applying clustering is shown in Fig.~\ref{fig:sec02_charge_sharing}, from flat-field measurements with an aluminum target at \SI{300}{\volt} bias voltage. Clustering was applied following the algorithm described above, respectively setting a primary and secondary discrimination threshold of $5\sigma$ and $3\sigma$, where $\sigma$ is the per pixel noise. Because of the narrow rectangular pixel geometry of this sensor, charge sharing is significantly higher in the short pixel pitch (vertical) direction. Therefore, a rectangular $2\times3$ pixels scanning window was defined. While the spectrum before clustering exhibits a pronounced tail toward lower energies due to charge sharing, after clustering the photon energy peak is reconstructed, becoming sharper and more symmetric. \par

For the described operational conditions, and for \SI{1.49}{\kilo\electronvolt} photons, the distribution of cluster sizes indicates that charge sharing occurs for \SI{\sim74}{\percent} of photon hits, with charge being shared among 2 pixels on \SI{\sim65}{\percent} of occasions, 3 pixels on \SI{\sim8}{\percent}, and 4 pixels on less than \SI{1}{\percent}. \par

Since charge sharing behavior can be modeled for a given detector geometry and set of operation parameters, it can also be exploited to estimate the location of the X-ray absorption with subpixel resolution. This is achieved through position interpolation algorithms, which analyze the distribution of charge among neighboring pixels to infer the most probable interaction point within the pixel cluster~\cite{Belau1983, Turchetta1993, Straulino2006}. \par

A straightforward approach for position interpolation is the Center of Gravity (CoG) method. In this technique, the estimated photon hit position is computed as the weighted average of the center positions of all pixels in the cluster, where the weights are given by the charge integrated in each pixel. While simple to implement and requiring no dedicated calibration, this method approximates the relationship between the photon hit position and the fraction of charge collected by each pixel as a linear one, which leads to a systematic error in the hit position determination~\cite{Turchetta1993}. \par

A more accurate method is based on the so-called $\eta$-function~\cite{Belau1983, Turchetta1993}, which accounts for the non-linear nature of charge sharing by modeling the charge distribution across neighboring pixels. The $\eta$-function is derived from flat-field measurements and can be computed using two-pixel clusters along a given dimension. In this configuration, the $\eta$ value is calculated as $\eta_\mathrm{2px} = T / (T + B)$, where $T$ and $B$ (i.e. \emph{top} and \emph{bottom} respectively) are the signals in two neighboring pixels along the energy dispersive dimension in a photon hit cluster. By collecting these statistics across the detector, a histogram of the $\eta$ values is built, characterizing the charge-sharing behavior. The interpolation function is then derived as the integral of this histogram and used as a look-up table to estimate the subpixel hit position~\cite{Ramilli2017}. A drawback of this method is that, since it relies on the empirically observed charge-sharing behavior, different $\eta$-function histograms and interpolation curves should be calculated to account for changing system conditions such as different photon energies, bias voltage or temperature. \par

\section{hRIXS spectrometer measurements}
\label{sec03}

Following the detector calibration and characterization measurements described in section~\ref{sec02}, the JUNGFRAU-iLGAD was mounted on the hRIXS spectrometer and tested with two reference solid samples: carbon tape (used for diagnostics) and bulk single-crystal cupric oxide, CuO (semiconducting sample used in this study for demonstration purposes). The main objectives were to evaluate the performance of the JUNGFRAU-iLGAD in comparison to the currently used Marana-X detector, particularly with respect to spatial resolution. Additionally, we aimed to demonstrate its capability for high frame rate (multi-kHz) RIXS measurements by exploiting its intra-train resolution capability. This is especially valuable for pump-probe experiments, where alternating pumped and unpumped X-ray pulses are recorded in consecutive memory cells, allowing direct comparison of the excited and ground state RIXS spectra to reveal fast transitions in the sample. \par

\subsection{Experimental setup}

The layout of the hRIXS spectrometer consists of only two elements: the grating and the detector. The optical design is variable line spacing (VLS) spherical, which allows for grazing focal planes~\cite{Schlappa2025}. Both, the grating chamber and detector chamber are movable, allowing for large adjustment of the lengths of the entrance and exit arms. In the grating chamber, two gratings are available currently: a high-transmission grating (1000 lines/mm) and a high-resolution grating (3000 lines/mm). The detector mounting flange is inclined at a fixed angle of \(\gamma=25^\circ \) in relation to the photon incidence direction, which enhances the effective spatial resolution by a factor of $1/\text{sin}(\gamma)\approx2.4$, but at the same time limits quantum efficiency due to the increase of the effective thickness of the entrance window. The JUNGFRAU-iLGAD was mounted in a 300~CF flange using an adapter flange. The samples were placed inside the XRD cryogenic sample environment chamber, where they were cooled to \SI{25}{\kelvin}. \par

Each \SI{10}{\hertz} pulse train delivered 304 FEL pulses at a repetition rate of \SI{1.1}{\mega\hertz}, with an average pulse energy of approximately \SI{900}{\micro\joule} before the monochromator. After \SI{50}{\percent} attenuation by the SASE3 gas attenuator (GATT)~\cite{Dommach2021}, the average pulse energy delivered to the sample was about \SI{94}{\nano\joule}, as measured by the X-ray Gas Monitors (XGMs)~\cite{Maltezopoulos2019}. The sample incidence angle of the FEL beam was \SI{42}{\degree} (from grazing) with a \SI{6.8}{\micro\meter} vertical spot size, and \SI{328}{\micro\meter} horizontal (values at $1/e^2$ assuming a beam with a Gaussian intensity profile), resulting in average X-ray fluence on sample of about \SI{1.8}{\milli\joule/\centi\meter\squared} as calculated from: \par

\begin{equation}
\text{F} = \frac{2E_p}{\pi w_{0,x}w_{0,y}}
\label{eq:fluence}
\end{equation}
where \(E_p\) is the pulse energy, and \(w_{0,x}\) and \(w_{0,y}\) are the horizontal and vertical spot sizes at $1/e^2$ intensity, respectively. \par

The X-ray polarization of the FEL was circular, with both positive and negative helicities (C$^+$ and C$^-$) tested during the beamtime. This polarization control is made possible by the APPLE-X undulators installed at the SASE3 beamline~\cite{Li2017}. The JUNGFRAU acquisition was synchronized to the FEL pulse train, with the 16 memory cells configured to integrate sequential exposure windows of \SI{18.95}{\micro\second} each, separated by a dead time of \SI{2.3}{\micro\second}, resulting in a burst acquisition rate of approximately \SI{47}{\kilo\hertz}. As a result, each memory cell integrates the signal produced by $304/16=19$ FEL pulses. \par

As in the measurements performed with the X-ray generator, the sensor was cooled using a chiller set to a bath temperature of \SI{-25}{\degreeCelsius}, with a continuous flow of silicone oil as coolant, and biased at \SI{300}{\volt} using a high voltage board mounted on an MPOD crate. \par

\subsection{Methods}
\label{sec03:methods}

\begin{figure}[!htb] %
\begin{center}
\includegraphics[width=0.8\textwidth]{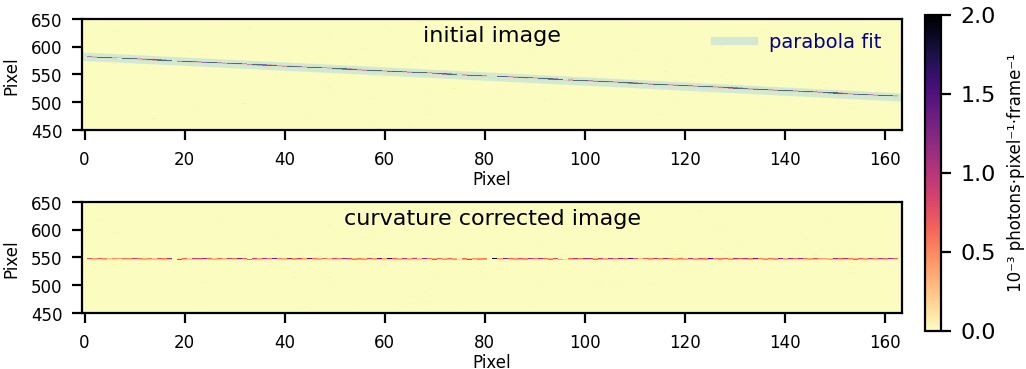}
\end{center}
\caption{Top: Average image showing a parabolic RIXS elastic line, and the corresponding parabola fit. Bottom: Same image after applying the curvature correction.}
\label{fig:sec03_curvature_correction}
\end{figure}

Before presenting the main results from the hRIXS measurements with the JUNGFRAU-iLGAD detector, we briefly outline the data analysis steps used to generate RIXS spectra from the detector output. The initial stages consist of the procedures commonly applied in hybrid pixel detectors, namely pedestal subtraction, gain correction, and bad pixel masking~\cite{Redford2018}. Following this, photon events are clustered and assigned subpixel vertical coordinates using the interpolation methods as described in section~\ref{subsec:charge_sharing}. Despite its limitations, the CoG interpolation method was applied instead of the $\eta$-function because an $\eta$-function was (also) not available for the Marana-X detector, which served as the spatial resolution benchmark. Therefore, to ensure a fair comparison between the two detectors, CoG interpolation was used for both. Following clustering and interpolation, the acquired frames in a given run are accumulated to obtain a two-dimensional image, since each individual frame typically contains only very few photons due to the low cross-section of RIXS. After this step, a sample elastic line with good SNR is obtained, such as the one represented in Fig.~\ref{fig:sec03_curvature_correction}-top, for the carbon tape (which provides strong non-resonant signal for the elastic line due to diffuse reflection). Because of the dispersive spherical grating, detected photons spread across a wide parabola in the detector plane. Before projecting the image to obtain the RIXS spectrum, this curvature must be corrected. This is achieved by fitting the photon distribution of the averaged 2D image with a parabola, and subsequently adjusting the vertical ($y$) coordinate of each photon hit using the retrieved parabola fit coefficients. The corrected vertical coordinates ($y'$) are then calculated as: \par

\begin{equation}
y' = y + (ax^2 + bx)
\label{eq:y_corr}
\end{equation}
\noindent
where $a$ and $b$ are the parabola coefficients, and $x$ the horizontal coordinate of the photon hit. As a result of this correction, a straight spectral line is obtained, as illustrated in Fig.~\ref{fig:sec03_curvature_correction}-bottom. \par

\subsection{Results and discussion}

\subsubsection{RIXS energy resolution}

\begin{figure}[!htb] %
\begin{center}
\includegraphics[width=1\textwidth]{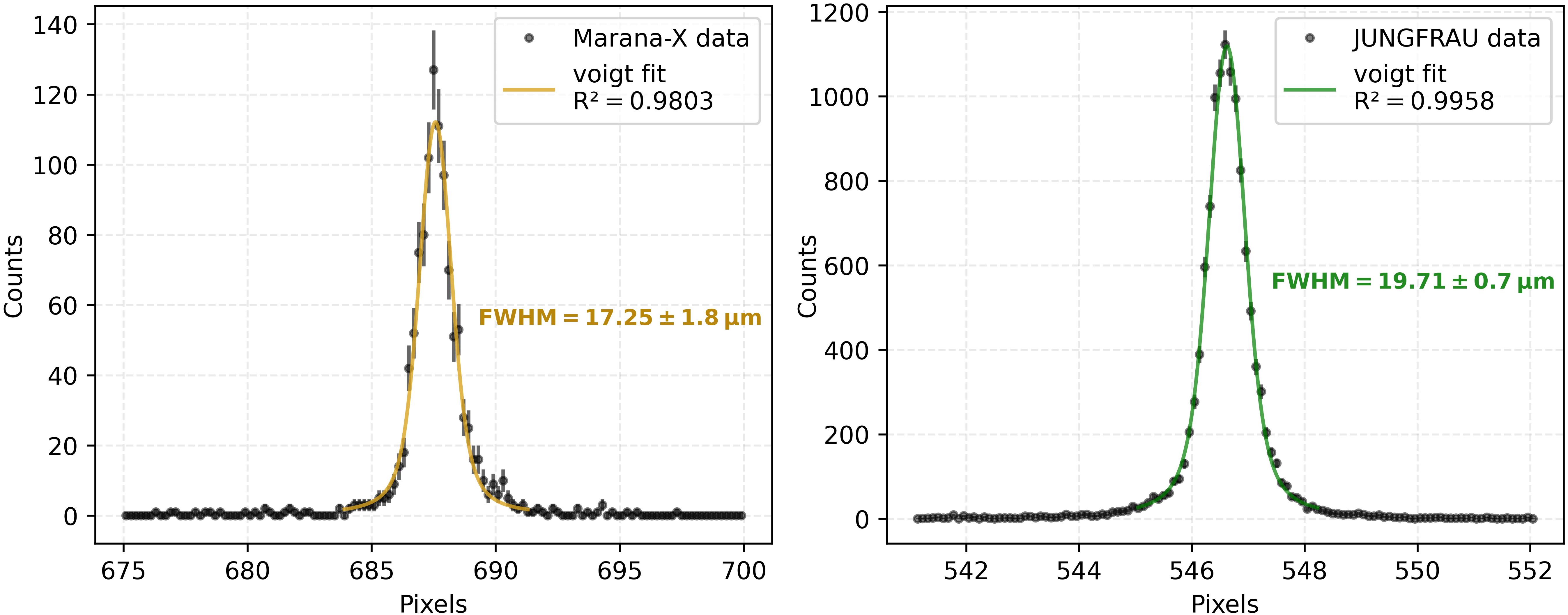}
\end{center}
\caption{RIXS elastic lines from \SI{928.7}{\electronvolt} photons on carbon tape, using the Marana-X (left) and JUNGFRAU-iLGAD (right) detectors, with the respective Voigt fits.}
\label{fig:sec03_resolution}
\end{figure}

In addition to evaluating the overall resolution of the hRIXS spectrometer with the JUNGFRAU-iLGAD detector, we also aimed to compare it with that achieved using the Marana-X detector. The total RIXS energy resolution is given by the convolution of the energy resolution of the SCS beamline and the energy resolution of the hRIXS spectrometer. Modifications on hRIXS have therefore a less direct impact on the total resolution. We do not have an independent measure to determine the energy resolution of the SCS beamline and rely on theoretical calculations for SASE3 monochromator. To this end, both measurements were conducted under identical experimental conditions. The spectrometer was set for 930 eV (Cu $L_3$-resonance edge) in back-scattering geometry with a scattering angle of $2\theta = 140^\circ$, and with the high-transmission spectrometer grating. This grating provides lower spectral dispersion than the high-resolution grating, resulting in a narrower RIXS elastic line, which is more suitable to characterize the detector spatial resolution. For both the Marana-X and the JUNGFRAU, the length of the exit arm was optimized to achieve the narrowest elastic line. The fitted spectra for both detectors are shown in Figure~\ref{fig:sec03_resolution}. \par

The obtained RIXS spectra are best fitted by a Voigt profile, a convolution of a Gaussian and a Lorentzian function, with a FWHM which can be approximated to~\cite{Kielkopf1973}: \par

\begin{equation}
f_{\mathrm{V}} \approx 0.5343\,f_{\mathrm{L}} + \sqrt{0.2169\,f_{\mathrm{L}}^2 + f_{\mathrm{G}}^2}
\end{equation}
\noindent
where \( f_{\mathrm{L}} \) and \( f_{\mathrm{G}} \) are respectively the FWHM of the Lorentzian and the Gaussian component. The resulting FWHM with the JUNGFRAU-iLGAD was $19.71\pm0.7~\si{\micro\meter}$, compared to $17.25\pm1.8~\si{\micro\meter}$ obtained with the Marana-X. \par

The higher uncertainty in the FWHM measured with the Marana-X detector is mostly justified by the lower photon statistics accumulated, due to a different FEL delivery pattern of 30 pulses per train, compared to 304 for the JUNGFRAU-iLGAD. Additionally, the Marana-X sensor has a width of \SI{\sim2}{\cm}, intercepting only about half of the photons emitted from the sample, compared to the JUNGFRAU-iLGAD sensor which has a total width of \SI{\sim4}{\cm}. These factors were, however, partially compensated by the higher quantum efficiency of the Marana-X (around \SI{99}{\percent} at the relevant photon energy~\cite{MaranaXwebsite}, compared to the \SI{\sim90}{\percent} of the tested JUNGFRAU-iLGAD), and the longer acquisition time adopted. \par

The \SI{\sim15}{\percent} difference in measured resolution can be primarily attributed to the smaller pixel size of the Marana-X detector (\SI{11}{\micro\meter}, compared to the \SI{25}{\micro\meter} of the JUNGFRAU-iLGAD in the spectroscopic direction). At this scale, the impact of pixel size is further accentuated by the prominence of charge sharing, which ultimately dictates how effectively the clustering and interpolation algorithms can perform. While charge sharing happens for virtually every photon hit in the Marana-X detector at \SI{930}{\electronvolt}, it is only observed for about half of the photon hits for the JUNGFRAU-iLGAD at the same photon energy conditions. Moreover, as previously discussed in section~\ref{subsec:charge_sharing}, the systematic uncertainty introduced by the CoG interpolation method increases when the charge cloud is small relative to the pixel size, further limiting the interpolation accuracy. \par

Additionally, the parabolic photon distribution on the detector plane introduces an uncertainty which is aggravated by the strong asymmetry of the JUNGFRAU-iLGAD strixels, as can be concluded from the $x$ dependency of the curvature correction equation \ref{eq:y_corr}. The liner coefficient \(b\) is superior to the quadratic coefficient \(a\) by a magnitude of $\sim\!10^{4}$, and therefore the latter can be neglected in equation~\ref{eq:y_corr}, while the former is given by $b=\tan(\theta)$ where $\theta$ is the angle between the spectral line and the horizontal axis of the detector. Deriving the error propagation for this equation, we find: \par

\begin{equation}
\sigma_{y'} = \sqrt{\sigma_{y}^2 + \tan^2(\theta) \, \sigma_x^2}
\label{eq:sigma_y_corr}
\end{equation}
\noindent
where $\sigma_{y}$ and $\sigma_{x}$ correspond to the spatial resolution in the vertical and horizontal directions, respectively. Because charge sharing is negligible on the horizontal direction, the center of the pixel is taken as estimate for each photon hit, and, attending to the \SI{225}{\micro\meter} pitch, can be assumed to correspond to \( \sigma_x=225/\sqrt{12}\) \si{\micro\meter} \cite{Turchetta1993}, following a uniform distribution ($\mathrm{FWHM}_x=\sigma_x$). On the vertical direction, in contrast, photon hit positions are derived from interpolation with the CoG method, and therefore follow a normal distribution ($\mathrm{FWHM}_y=2\sqrt{2\ln2}\,\sigma_y \approx 2.355 \sigma_y$). With these considerations, we can re-write equation \ref{eq:sigma_y_corr} in terms of its FWHM: \par

\begin{equation}
\mathrm{FWHM}_{y'} = \sqrt{\mathrm{FWHM}_{y}^2 + (2.355\,\tan(\theta)\,\mathrm{FWHM}_x)^2}
\end{equation}

In this experimental setup, the angle of the photon distribution on the detector plane was approximately $\theta=\ang{2.8}$, the contribution of this effect to the overall resolution amounts to \SI{7.5}{\micro\metre}, corresponding to about 38\% of the measured \SI{19.7}{\micro\metre} resolution. To mitigate this contribution in future measurements, the detector tilt should remain below \SI{1}{\degree}. \par

It is also worth noting that, in contrast to the Marana-X detector which has been in user operation for some time, the analysis tools for the JUNGFRAU detector were not developed at a similar high level and the presented test measurements had to be accomplished within limited time. This meant that some aspects of the JUNGFRAU-iLGAD data processing, such as bad pixel masking and interpolation, could only be properly optimized using data acquired during the beamtime. As a result, only a preliminary version of the processing pipeline was available to provide rapid feedback to guide the measurements, particularly in the determination of the spectrometer focal plane. Consequently, it is likely that the scanning of the detector position along the exit arm of the spectrometer could have been more precise, suggesting that there is room for improvement of the achieved spatial resolution. \par

Overall, the measured energy resolutions are comparable for both detectors. This indicates that the current resolution limit is primarily determined by other factors, rather than by the detector intrinsic resolution. A more common metric to estimate the spectrometer resolution is the resolving power, which is defined as the ratio between a given photon energy ($E$) and the smallest difference in energy ($\Delta E$) that still allows to resolve two separate peaks: \par

\begin{equation}
R = \frac{E}{\Delta E}
\end{equation}

To evaluate the resolving power measured with the JUNGFRAU-iLGAD detector, the grating was changed to the high-resolution one, which provides higher energy dispersion, enabling a better separation of closely spaced energy features on the detector. \par

\begin{figure}[!htb] %
\begin{center}
\includegraphics[width=0.5\textwidth]{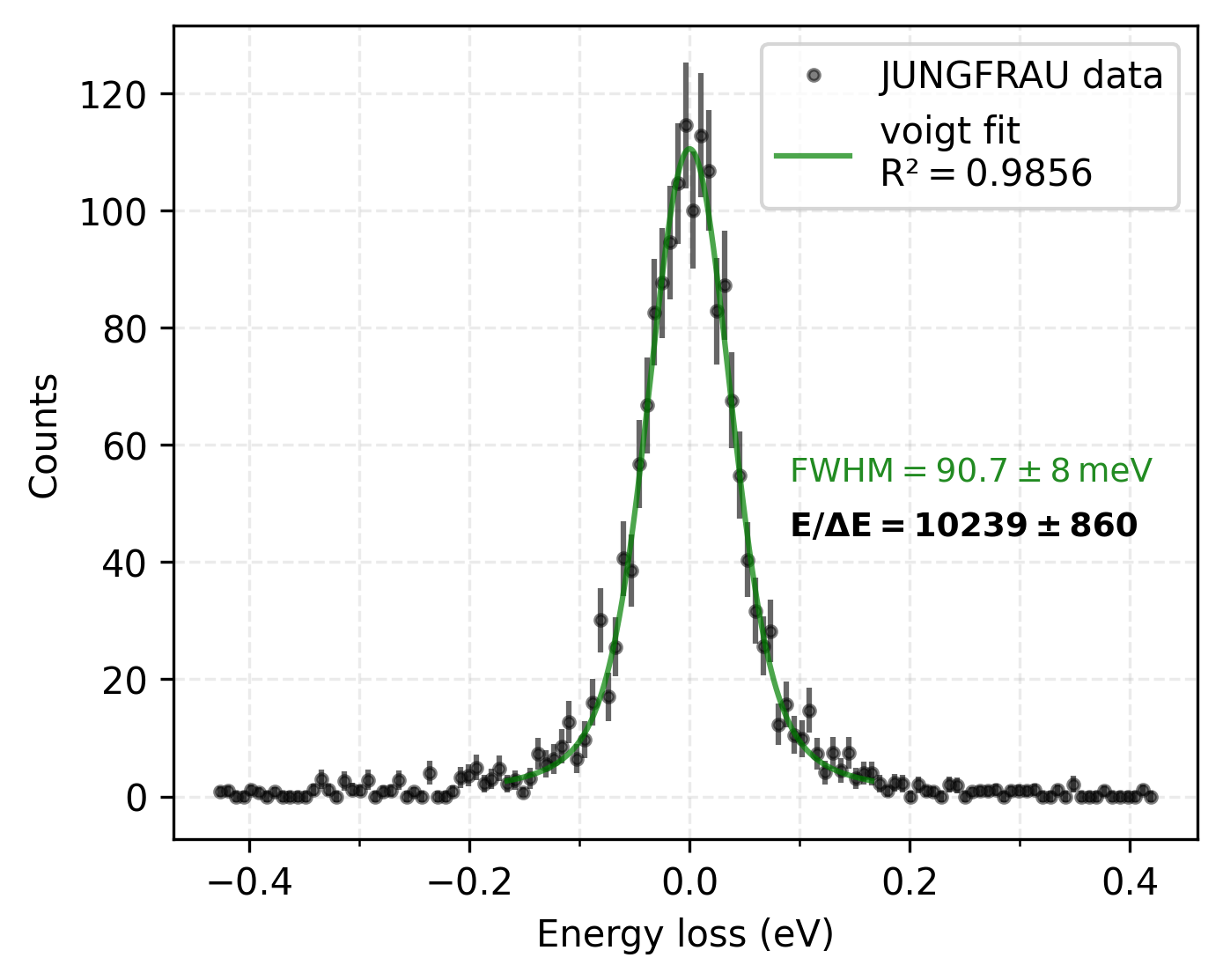}
\end{center}
\caption{Resolving power measured at \SI{928.5}{\electronvolt} with the high-resolution grating.}
\label{fig:sec03_resolving_power}
\end{figure}

The parameter $\Delta E$ corresponds to the total energy resolving power, and can be quantified by the FWHM of the RIXS line after converting from pixel position to energy.  This calibration was performed by acquiring multiple spectra from the carbon tape sample while scanning the incident photon energy across the \SI{924.5}{\electronvolt} to \SI{933.5}{\electronvolt} range. The resulting linear relationship between photon energy and the centroid of the fitted RIXS spectrum (in pixel units) was used to convert from pixel position to energy units. A longer acquisition at a reference energy of \SI{928.5}{\electronvolt} was then used to access the resolving power, with the result plotted in Fig.~\ref{fig:sec03_resolving_power}. A resolving power above 10,000 was measured for the referred photon energy. This result is consistent with previous measurements performed under similar conditions at \SI{930}{\electronvolt}~\cite{Schlappa2025} and is also close to the theoretical value for the SASE3 monochromator~\cite{Gerasimova2022}. \par

\subsubsection{XFEL effect on sample RIXS signal}

In order to assess potential limitations of measurements with the hRIXS spectrometer, it is important to evaluate whether the FEL or pump-probe laser (PPL) pulses can induce cumulative effects in the sample that could compromise the interpretation of the RIXS spectra. This would mean that within a FEL train, the action of earlier pulses influences the sample behavior probed by later pulses due to, e.g., accumulative heating, meaning that the successive pulses across the train should not be considered statistically independent. This would, in turn, lead to inaccuracies when averaging over the train pulses. \par

\begin{figure}[!htb] %
\begin{center}
\includegraphics[width=0.7\textwidth]{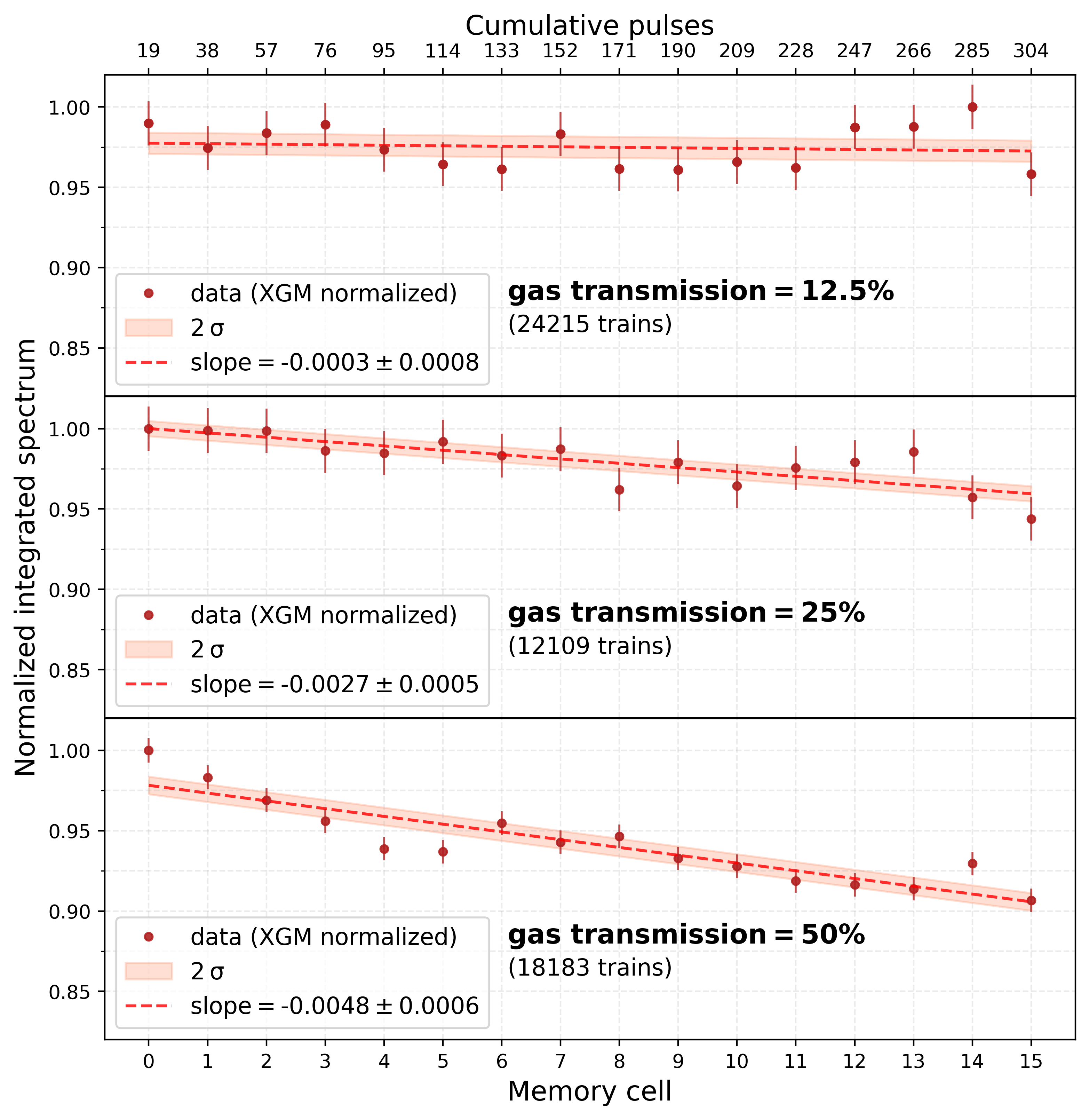}
\end{center}
\caption{Integrated X-ray emission from CuO sample per memory cell for different FEL intensities at \SI{928.5}{\electronvolt} incident photon energy, normalized for the XGM recorded pulse intensities. Signal reduction across memory cells is observed for higher transmissions, indicating FEL-induced effects.} \label{fig:sec03_FEL_effect}
\end{figure}

The newly provided ability to resolve separated pulses within an XFEL train enables the investigation of intra-train cumulative effects, providing the conditions to characterize the extent to which FEL-induced changes in the sample may condition high repetition rate RIXS measurements. Profiting from this, a dedicated set of measurements was carried out without the PPL, focusing only on effects induced by the X-rays. Different FEL intensities were delivered to the CuO sample at \SI{928.5}{\electronvolt} photon energy by adjusting the transmission of the upstream GATT. The results, illustrated in Fig.~\ref{fig:sec03_FEL_effect} obtained by integrating the whole RIXS spectrum region for the different gas transmissions, show that a decrease in the number of detected photons emitted by the sample along successive memory cells was observed, over a time range of \SI{340}{\micro\second}. It can also be seen that this effect is negligible for the lowest gas transmission of \SI{12.5}{\percent}, it is moderate at \SI{25}{\percent} gas transmission, and it becomes evident at the highest tested transmission of \SI{50}{\percent}. These gas transmission percentages correspond to an average peak fluence on the sample of \SI{0.45}{\milli\joule/\centi\meter\squared}, \SI{0.9}{\milli\joule/\centi\meter\squared} and \SI{1.8}{\milli\joule/\centi\meter\squared}, respectively. The plotted data points correspond to the total intensity integrated by each of the memory cells averaged over multiple trains. To account for the intrinsic pulse-to-pulse intensity fluctuations of the SASE FEL, the data was normalized according to the pulse-resolved intensity values recorded by the XGMs. The increasing magnitude of this trend, as quantified by the slope of a linear fit to the data presented in the plots, supports its attribution to FEL-induced effects. If the material properties do not recover between pulses in a single train, the loss of signal would also be expected to be noticeable in X-ray absorption spectroscopy (XAS). In this case, the XAS signal (e.g. the total fluorescence yield) could be measured before performing RIXS experiments in order to determine the limit of fluence and number of pulses per train that the sample can withstand without significant FEL-induced effects. The JUNGFRAU detector enables ``on the fly" monitoring and detection of possible accumulated effects, directly from the recorded RIXS spectra. \par 

The decreasing trend in sample RIXS signal is not observed when averaging a fixed memory cell over multiple trains, suggesting that the sample recovers from the FEL-induced effects between consecutive trains. \par

\subsubsection{Pump-probe RIXS studies}

We conducted RIXS measurements from CuO, after exciting the sample with the pump-probe laser (PPL). The PPL delivered \SI{50}{\femto\second} pulses of \SI{800}{\nano\meter} wavelength with 
a peak fluence of \SI{10.3}{\milli\joule/\centi\meter\squared} (Gaussian beam profile with vertical (horizontal) spot size of \SI{320}{\micro\meter} (\SI{550}{\micro\meter})). It was set to the same XFEL repetition rate of \SI{1.1}{\mega\hertz}, with the XFEL pulses delayed by \SI{1}{\pico\second} with respect to PPL. The FEL was configured with positive circular polarization and gas transmission of \SI{25}{\percent}. \par

To enable direct comparison between pumped and unpumped conditions, the PPL pulses were delivered only to alternating memory cells of the JUNGFRAU-iLGAD detector, i.e., even-numbered cells recorded RIXS data without PPL excitation (unpumped data), while odd-numbered cells recorded RIXS data while PPL pulses were delivered (pumped data). The unpumped data serves as the baseline for the equilibrium state of the sample, when no laser-induced excitations take place, thus providing a reference for pumped data normalization. This scheme provides opportunity to investigate the effect of heat accumulation in the sample during the train, which ideally should be minimized. \par

\begin{figure}[!htb] %
\begin{center}
\includegraphics[width=0.7\textwidth]{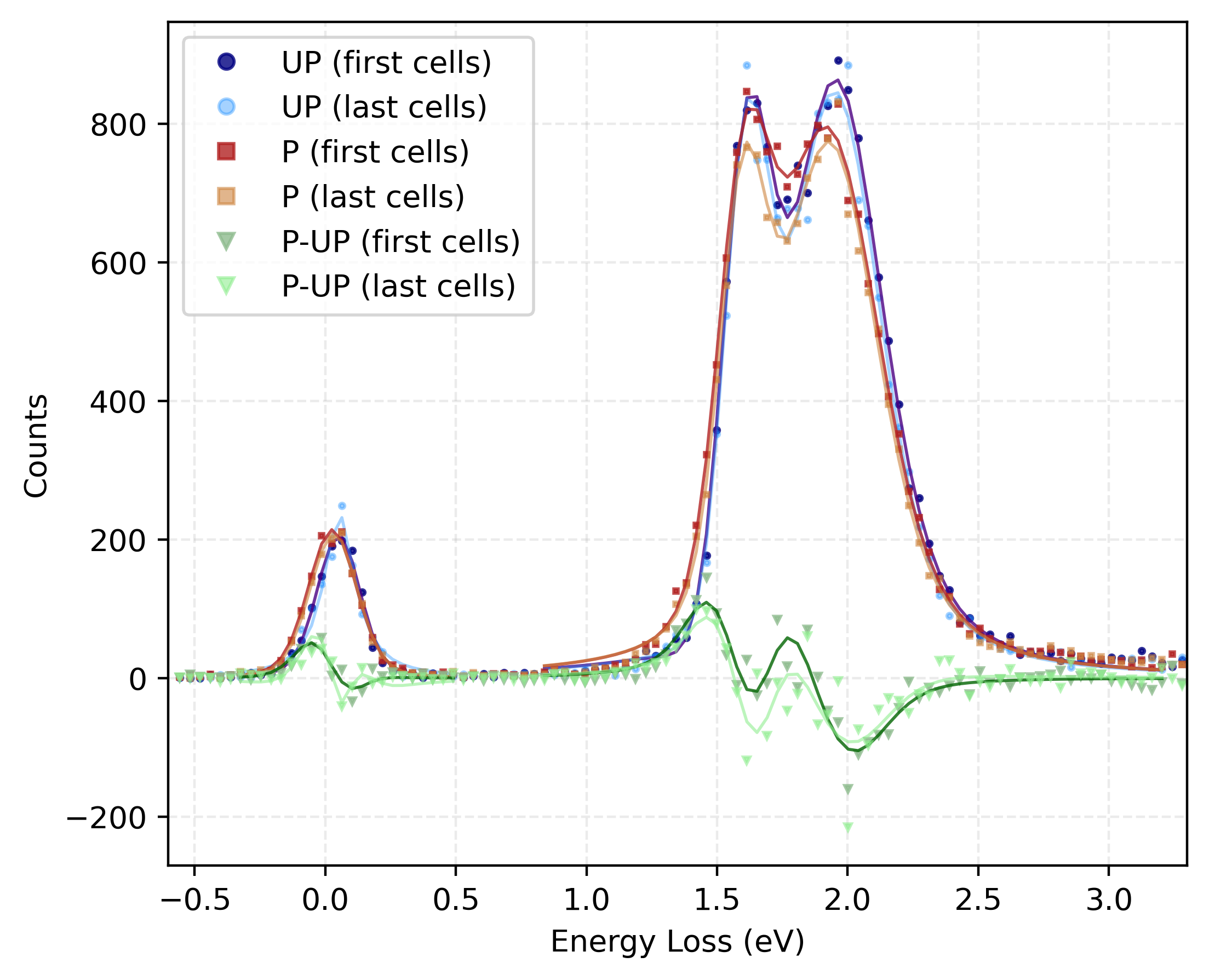}
\end{center}
\caption{RIXS spectra of a CuO sample at \SI{928.5}{\electronvolt} photon energy acquired 1 ps after PPL excitation (P) and without (UP), as well as their difference spectra (P-UP). The solid lines are fits that serve as guide to the eye. Each set was obtained from first half (memory cells 0–8) and second half of the pulse train (memory cells 9–15).}
\label{fig:sec03_PPL_effect}
\end{figure}

An example of RIXS spectra acquired under the described conditions at \SI{928.5}{\electronvolt} incident photon energy is shown in Fig.~\ref{fig:sec03_PPL_effect}. Both, the pumped and unpumped data were divided into two subsets: the first half of the pulse train (memory cells 0–8) and the second half (memory cells 9–15). This separation allows to evaluate potential changes in the sample response over the duration of the train. The spectra for the pumped and unpumped conditions were obtained by averaging the corresponding frames over their respective memory cells. \par

For all datasets, two main features can be distinguished in the RIXS spectra: a quasi-elastic line (elastic and inelastic signal merging into one peak) close to zero energy loss, and a double-peak feature between 1.2 and 2.5 eV energy loss which corresponds to orbital $d-d$ excitations at the Cu sites~\cite{Ghiringhelli2009}. The data shows clear differences between the pumped and unpumped spectra, indicating the presence of fast laser-induced effects in CuO. Looking at slow changes across the pulse train, while the unpumped spectra remain mostly stable, there is a small change in the pumped spectra around \SI{1.8}{\electronvolt} between the first and second halves of the train. This can be interpreted as a slight sample heating during the pump sequence, due to energy deposition from the optical laser, possibly combined with X-ray heating. \par

While a more thorough interpretation of these laser-induced effects is beyond the scope of this work, the results demonstrate the capability of the JUNGFRAU–iLGAD detector to resolve such accumulated effects and serves as a proof of concept for pump-probe RIXS. By scanning a range of time delays between the PPL and FEL pulses, the dynamic evolution of the system is accessed, enabling time-resolved RIXS with improved reliability. \par

\section{Conclusions and future work}
\label{sec04}

The results presented in this work mark a significant step towards unlocking the full potential of the hRIXS spectrometer at the SCS instrument of the European XFEL, particularly by enabling the investigation of fast sample dynamics that occur within an XFEL pulse train. Furthermore, we demonstrated the suitability of a JUNGFRAU detector equipped with an iLGAD sensor to be used in time-resolved RIXS experiments, offering not only high spatial resolution and good signal-to-noise ratio, but also high repetition rates and a large sensitive area, establishing it as a promising candidate to overcome the limitations imposed by the commonly used alternatives based on CCD or CMOS technologies. \par

A spatial resolution of $19.71\pm0.7~\si{\micro\meter}$ was measured at \SI{928.5}{\electronvolt} using the JUNGFRAU-iLGAD, comparable to that of Marana-X camera which has been the standard detector used with the hRIXS spectrometer. A resolving power exceeding 10,000 was also achieved. It is important to notice that the obtained resolution is currently limited by the energy resolution of SASE3 optics, rather than by the intrinsic detector resolution~\cite{Gerasimova2022,Schlappa2025}. Additional work is ongoing to estimate the intrinsic spatial resolution of the same JUNGFRAU-iLGAD prototype used in the hRIXS measurements, through dedicated knife-edge measurements using the X-ray generator setup described in section~\ref{sec02}. \par

Profiting from the burst mode acquisition of the JUNGFRAU, RIXS spectra from reference samples were acquired at an unprecedented frame rate of \SI{\sim47}{\kilo\hertz}. Pump-probe measurements under realistic experimental conditions were performed, with clearly visible differences between the pumped and unpumped spectra in CuO, reinforcing the robustness of the data normalization achieved by alternating the acquisition of pumped and unpumped pulses across consecutive memory cells of the detector. \par

The high frame-rate achieved also allowed to reveal a decrease in sample RIXS signal across the XFEL pulse train from the (semiconducting) CuO sample. This effect was shown to increase with FEL intensity, and a signal reduction of roughly \SI{10}{\percent} was observed between the first and last memory cells of the JUNGFRAU detector for 928.5 eV pulses depositing an average peak fluence of \SI{1.8}{\milli\joule/\centi\meter\squared}. This observation suggests that intense FEL pulses might induce reversible accumulated changes in samples. The identification of such phenomena is crucial in high frame rate RIXS applications, in order to avoid that the sample response to the multiple FEL pulses across the train is biased by other effects other than the optical laser excitation pulse. Further studies to better characterize this effect could be achieved by scanning the gas transmission to cover an extended intensity range with finer granularity. It should be mentioned that the thresholds for accumulated changes will vary for different sample systems. \par

Efforts are also ongoing to improve the quantum efficiency of iLGAD and standard silicon sensors by equipping them with thinner entrance windows. For recent iterations, quantum efficiencies of up to \SI{82}{\percent} at \SI{500}{\electronvolt} and \SI{62}{\percent} at \SI{250}{\electronvolt} have been reported~\cite{Carulla2024}, and the newest sensor batch with process variations aiming for yet higher quantum efficiencies is currently being characterized, with results expected to be published in the near future. These new advancements improving quantum efficiency are particularly relevant for the hRIXS case when taking into consideration its $25^\circ$ detector mounting angle inclination, which makes for an increase of the perceived path length of the incoming photons by a factor of $1/\sin(25^\circ) \approx 2.4$. The expected improvement of quantum efficiency should allow to extend the JUNGFRAU-iLGAD sensitivity to energies down to \SI{500}{\electronvolt} and below, entering the so-called ``water window" and expanding the range of studies that could profit from the hRIXS spectrometer capabilities. \par

Finally, different iLGAD sensor geometries with increased segmentation factor have been successfully manufactured and bump-bonded with JUNGFRAU chips, providing pitches of \SI{18.75}{\micro\meter} and \SI{15}{\micro\meter}. Such geometries should provide a significant improvement of spatial resolution due the increased charge sharing. Additionally, preliminary tests showed that the characterization of the $\eta$-function in similar operation conditions and photon energy to hRIXS measurements should be possible by using a zero order grating, effectively defocusing the spectrometer to cover the whole sensor surface and achieve a relatively flat illumination. This would allow to use the $\eta$-function algorithm without the need of a separate flat-field measurement campaign, which would in term allow to eliminate the systematic interpolation uncertainty inherent to the CoG method, contributing to an improvement in spatial resolution. \par

\begin{acknowledgements}
We acknowledge the European XFEL in Schenefeld, Germany, for provision of X-ray free-electron laser beamtime at SCS SASE3 under proposal number 8716, and would like to thank the staff for their assistance.
\end{acknowledgements}

\begin{funding}
This research was funded in part by the Swiss National Science Foundation under grant number PZ00P2\_223377. M. G. was supported by the Swiss National Science Foundations under grant number 10.000.807. For the purpose of open access, a CC BY public copyright license is applied to any author accepted manuscript (AAM) version arising from this submission.  
\end{funding}

\DataAvailability{Data recorded for the experiment at the European XFEL are available at doi: {10.22003/XFEL.EU-DATA-008716-00.}}

\bibliography{iucr} 

\end{document}